\documentclass[12pt]{iopart}
\usepackage{amstext}
\usepackage{iopams}
\usepackage{graphicx}

\begin{document}

\title[Population oscillations in Lotka--Volterra models]
      {Population oscillations in spatial stochastic Lotka--Volterra models: 
       A field-theoretic perturbational analysis}

\author{Uwe C. T\"auber}

\address{Department of Physics, Virginia Tech, Blacksburg, VA 24061-0435, USA}

\ead{tauber@vt.edu}


\begin{abstract}
Field theory tools are applied to analytically study fluctuation and 
correlation effects in spatially extended stochastic predator-prey systems.
In the mean-field rate equation approximation, the classic Lotka--Volterra 
model is characterized by neutral cycles in phase space, describing undamped 
oscillations for both predator and prey populations.
In contrast, Monte Carlo simulations for stochastic two-species predator-prey 
reaction systems on regular lattices display complex spatio-temporal structures
associated with persistent erratic population oscillations.
The Doi--Peliti path integral representation of the master equation for 
stochastic particle interaction models is utilized to arrive at a field theory 
action for spatial Lotka--Volterra models in the continuum limit.
In the species coexistence phase, a perturbation expansion with respect to the 
nonlinear predation rate is employed to demonstrate that spatial degrees of
freedom and stochastic noise induce instabilities toward structure formation,
and to compute the fluctuation corrections for the oscillation frequency and 
diffusion coefficient.
The drastic downward renormalization of the frequency and the enhanced 
diffusivity are in excellent qualitative agreement with Monte Carlo simulation 
data.
\end{abstract}

\submitto{\JPA --- \today}

\section{Introduction}
\label{sec:Introduction}

In the past two decades or so, analytical and computational tools developed in
statistical physics have been quite successfully applied to mathematical 
problems in ecology and population dynamics, with the overall goal to arrive at
a quantitative understanding of the emergence of biodiversity in nature 
\cite{May73}--\cite{Murray02}. 
The typical physics approach to complex dynamical systems is of course to first
consider perhaps oversimplified idealized models that however are designed to
hopefully capture the essential phenomenology.
A considerable part of the mathematical biology literature largely addresses 
coupled deterministic equations of motion for interacting population species that 
are ultimately based on mean-field type rate equation approximations, whereas
leaving aside some of the biological complexity provides the opportunity to 
consistently incorporate stochastic fluctuations and spatio-temporal correlations, 
whose crucial importance has long been recognized in the field \cite{Durrett99}.

This paper addresses predator-prey competition models that are defined via 
reaction-diffusion systems on a regular $d$-dimensional lattice, and whose rate
equations in the well-mixed mean-field limit reduce to the two coupled 
ordinary differential equations originally introduced independently by Lotka 
\cite{Lotka20} and Volterra \cite{Volterra26} nearly a century ago.
These stochastic spatial predator-prey models have served as paradigmatic 
examples for the emergence of cooperative steady states in the dynamics of two 
competing populations \cite{Matsuda92}--\cite{Boccara94}
(see also Ref.~\cite{Mobilia07} for a fairly recent overview). 
The deterministic Lotka--Volterra rate equation model is characterized by a 
neutral cycle in phase space, describing regular undamped nonlinear population
oscillations with the unrealistic feature that both predator and prey 
population densities invariably return to their initial values.
In contrast, computer simulations of sufficiently large stochastic 
Lotka--Volterra systems yield long-lived erratic population oscillations 
\cite{Provata99}--\cite{Kowalik02}, 
whose persistence can be understood through a resonant stochastic amplification
mechanism \cite{McKane05} that drastically extends the transient time interval 
before any finite system ultimately reaches its absorbing stationary state, 
where the predator population becomes extinct \cite{Parker09, Dobrinevski12}.
In spatially extended systems, the mean-field Lotka--Volterra reaction-diffusion 
equations allow for traveling wave solutions \cite{Dunbar83}--\cite{Aguiar04}. 
In the corresponding stochastic lattice realizations, these regular wave crests
become spreading activity fronts \cite{Movies} that further enhance both
populations' life span, and furthermore induce short-ranged but significant 
positive correlations between representatives of either species, and 
anti-correlations between the predator and prey populations
\cite{Mobilia07, Washenberger07}.
Over the past years, we have investigated various different variants of such
two-species stochastic spatial Lotka--Volterra models for competing 
predator-prey populations, and found these intriguing spatio-temporal 
structures to be remarkably stable with respect to modifications of the 
detailed microscopic interaction rules \cite{Washenberger07, Mobilia06}.
Even in the presence of quenched spatial disorder in the reaction rates, the
qualitative features of spatial stochastic Lotka--Volterra models remain 
unchanged, although quite remarkably both the predator and prey populations 
benefit from such environmental variability \cite{Dobramysl08}.

Many qualitative features of stochastic spatial predator-prey systems are 
adequately captured by the associated coupled mean-field rate equations, 
augmented with diffusive spreading.
However, one observes strikingly strong quantitative renormalizations of, e.g.,
the characteristic population oscillation frequency, whose numerically 
determined values in various systems were found to be reduced as compared with
the (linearized) rate equation predictions by factors in the range 
$2 \ldots 6$, depending on the reaction rates, both in the presence and absence
of site occupation number restrictions for the predator and prey populations 
\cite{Mobilia07, Washenberger07}.
In addition, the neutral cycle oscillations in the original Lotka--Volterra 
rate equations are undamped; in contrast, when a finite carrying capacity for 
the prey species is imposed, the system relaxes to a stable coexistence fixed 
point.
However, starting from random initial states, Monte Carlo simulations for the 
corresponding stochastic lattice models yield initially damped population 
oscillations in the coexistence phase, even in the absence of any restrictions 
on the site occupation numbers, i.e., for infinite local carrying capacities.
These are associated with striking spatio-temporal structures, namely spreading
waves of prey closely followed by predators.
In our Monte Carlo simulations, we measured the front speed to be markedly 
enhanced with respect to the mean-field prediction \cite{Dobramysl08}.

The aim of this present paper is to provide a qualitative and even 
semi-quantitative explanation for these intriguing observations.
The Doi--Peliti coherent-state path integral representation of the master 
equation for stochastic interacting particle systems 
\cite{Doi76}--\cite{Andreanov06} (for recent reviews, see 
Refs.~\cite{Mattis98, Tauber05}), augmented with a means to incorporate 
restricted site occupation numbers \cite{Wijland01}, will be employed to gain a 
comprehensive understanding of fluctuation and correlation effects in the
thermodynamic limit of stochastic spatial predator-prey models.
More specifically, a perturbative loop expansion to first order in the 
nonlinear predation rate will be constructed; it will allow us to demonstrate
the instability of the spatial stochastic system against dynamic structure 
formation, and enable the computation of the fluctuation-induced 
renormalizations of the population oscillation frequency and diffusion 
coefficient \cite{Tauber11}.

This very same formalism was already utilized in Ref.~\cite{Mobilia07} to 
demonstrate with the aid of renormalization group arguments that the effective 
critical field theory in the vicinity of the predator extinction threshold that
emerges at low predation rate for finite prey carrying capacity can be mapped 
onto Reggeon field theory which encapsulates the universal scaling behavior of 
critical directed percolation clusters \cite{Obukhov80}--\cite{Janssen05}. 
Indeed, since the predator extinction transition represents a continuous 
nonequilibrium phase transition from an active stationary to an inactive, 
absorbing state (in the absence of any conserved quantities and quenched 
disorder), one would expect it to be governed by the prominent directed 
percolation universality class \cite{Janssen81}--\cite{Odor04}.
There exists now ample numerical evidence that the critical exponent values at
or near the predator extinction transition in spatially extended 
Lotka--Volterra systems are in fact those of directed percolation 
\cite{Tome94}--\cite{Kowalik02}. 

The present work also complements and transcends the treatment in 
Ref.~\cite{Butler09} where the same mathematical framework was utilized as a
starting point for a van Kampen system size expansion, demonstrating on the
Gaussian fluctuation level the persistence of population oscillations in the 
species coexistence phase of stochastic lattice Lotka--Volterra models, thus 
generalizing the zero-dimensional analysis in Ref.~\cite{McKane05} to spatially 
extended systems.
Here, the fluctuation-induced renormalizations of the characteristic population
oscillation frequency, damping, and diffusivity will be computed to first order in 
a perturbative expansion with respect to the nonlinear predation rate.
It should be noted that in contrast with the powerful van Kampen system size
expansion, the Doi--Peliti formalism captures fluctuations and intrinsic reaction
noise in the thermodynamic limit, and has been successfully applied to systems
governed by strong correlations for which a simple Kramers--Moyal expansion 
and Fokker--Planck truncation fails (see, e.g., Ref.~\cite{Tauber05}).
The field-theoretic approach has also been employed as an efficient route to 
construct an expansion about the thermodynamic limit in spatially extended
predator-prey systems in Refs.~\cite{Goldenfeld09, Butler11}.

This paper is structured as follows:
The following section begins with a concise review of the properties of the 
Lotka--Volterra mean-field rate equations, including the modifications induced 
by a finite prey carrying capacity, and some crucial features observed in Monte
Carlo simulations for stochastic two-species predator-prey models on a regular 
lattice.
Next the construction of the Doi--Peliti path integral representation is 
explained, and its utility demonstrated by a brief summary of the crucial steps
that allow a mapping of the Lotka--Volterra system with finite local prey 
carrying capacity near the predator extinction threshold onto Reggeon field 
theory that governs the directed percolation universality class.
Subsequently, this formalism is employed to construct a systematic perturbation
expansion with respect to the nonlinear predation rate in the species 
coexistence phase.
We then establish the presence of structure formation instabilities, and 
proceed to compute the renormalized population oscillation frequency and 
diffusivity to one-loop order, and compare our results with simulation data.
The conclusion summarizes these novel results, and gives an outlook to future 
investigations.
Two appendices provide additional technical details and an integral table.

\section{Stochastic lattice Lotka--Volterra models}

We begin by first defining the stochastic interacting particle model under
consideration through a set of coupled irreversible `chemical' reactions, and 
then provide a summary of its basic features as obtained in the mean-field rate
equation approximation.
Next we discuss the crucial numerical observations from the extensive 
literature for stochastic spatially extended two-species predator-prey systems.

\subsection{Model variants and mean-field description}

We consider a system comprised of two distinct particle species that propagate
diffusively with continuum diffusion constants $D_{A/B}$ and undergo the 
following stochastic reactions:
\begin{eqnarray}
  &A \to \emptyset  \quad &{\rm with \ rate} \ \mu , \nonumber \\
  &A + B \to A + A  \quad &{\rm with \ rate} \ \lambda' , 
\label{lvreac} \\
  &B \to B + B \quad &{\rm with \ rate} \ \sigma . \nonumber
\end{eqnarray}
The `predators' $A$ decay or die spontaneously at rate $\mu > 0$, whereas the 
`prey' $B$ produce offspring with rate $\sigma > 0$.
In the absence of the binary `predation' interaction with rate $\lambda$, the 
uncoupled first-order processes would naturally lead to predator extinction 
$a(t) = a(0) \, e^{- \mu t}$, and Malthusian prey population explosion
$b(t) = b(0) \, e^{\sigma t}$; here $a(t)$ and $b(t)$ respectively indicate the
$A (B)$ concentrations or population densities.
The predation reaction constitutes a nonlinear interaction that simultaneously
controls the prey particle number and allows the predators to multiply, thus 
opening the possibility of species coexistence through competition.

In the simplest spatial realization of this stochastic reaction-diffusion 
model, both particle species are represented by unbiased random walkers on a 
$d$-dimensional hypercubic lattice (with lattice constant $a_0$), and one 
allows an arbitrary number of particles per lattice site (see 
Ref.~\cite{Washenberger07}).
The reactions (\ref{lvreac}) can then all be implemented strictly on-site: 
Offspring particles are placed on the same lattice point as their parents, and 
the predation reaction happens only if an $A$ and a $B$ particle meet on the 
same lattice site.
If one further assumes the populations to remain well mixed, and consequently 
ignores both spatial fluctuations and correlations, the coupled reactions 
(\ref{lvreac}) can approximately be described through the associated mean-field
rate equations for spatially homogeneous concentrations 
$a(t) = \langle a({\vec x},t) \rangle$, $b(t) = \langle b({\vec x},t) \rangle$,
where $a({\vec x},t)$ and $b({\vec x},t)$ respectively denote the local 
predator and prey densities.
With $\lambda = a_0^d \, \lambda'$, this leads to the two classical 
Lotka--Volterra coupled ordinary nonlinear differential equations 
\cite{Murray02}:
\begin{equation}
  \dot{a}(t) = \lambda \, a(t) \, b(t) - \mu \, a(t) \ , \quad
  \dot{b}(t) = \sigma \, b(t) - \lambda \, a(t) \, b(t) \ .
\label{lvreqa}
\end{equation}

The rate equations (\ref{lvreqa}) display three stationary states $(a_s,b_s)$:
(i) the empty absorbing state (total population extinction) $(0,0)$, which is 
obviously linearly unstable if $\sigma > 0$; (ii) an absorbing state wherein 
the predators go extinct and the prey population diverges $(0,\infty)$, which 
for $\lambda > 0$ is also linearly unstable; and (iii) a species coexistence 
state $(a_u = \sigma/\lambda , b_u = \mu/\lambda)$, which represents a 
marginally stable fixed point with purely imaginary eigenvalues 
$\pm i \omega_0$ of the associated Jacobian stability matrix, with the (linear)
oscillation frequency $\omega_0 = 2 \pi \, f_0 = \sqrt{\mu \sigma}$ about the 
center fixed point $(a_u,b_u)$.
In the full nonlinear ordinary differential equation system (\ref{lvreqa}), 
the phase space trajectories are determined by 
$d a / d b = [a \, (\lambda \, b - \mu)] / [b \, (\sigma - \lambda \, a)]$, 
for which one easily identifies the conserved first integral 
$K(t) = \lambda [a(t) + b(t)] - \sigma \ln a(t) - \mu \ln b(t) = K(0)$.
As a consequence, the solutions of the deterministic Lotka--Volterra rate
equations form closed orbits in phase space that describe regular periodic 
nonlinear population oscillations whose amplitudes and phases are fixed by the
initial configuration.
Naturally, the neutral cycles of the coupled mean-field rate equations  
(\ref{lvreqa}) that appear independent of the set rates and take the system
precisely back to its initial configuration after one period represent 
biologically unrealistic features, and are moreover indicative of the 
fundamental instability of this deterministic mathematical model with respect 
to even slight modifications \cite{Murray02}.

One important example of such an alteration that aims at rendering the 
Lotka--Volterra system more relevant to ecology is to introduce a finite 
carrying capacity (maximum total particle density) $\rho > 0$ that limits the 
prey population growth \cite{Murray02}.
It can be interpreted as originating from, e.g., limited food resources for
the prey species.
Within the mean-field rate equation framework, the second differential equation
in (\ref{lvreqa}) then becomes replaced with
\begin{equation}
  \dot{b}(t) = 
  \sigma \, b(t) \left[ 1 - b(t) / \rho \right] - \lambda \, a(t) \, b(t) \ . 
\label{lvreqc}
\end{equation}
Again, one finds three stationary states in this restricted Lotka--Volterra 
model: (i) total extinction $(0,0)$; (ii') predator extinction and prey 
saturation $(0,\rho)$, which becomes linearly stable for sufficiently small
predation rates $\lambda < \lambda_c = \mu / \rho$; (iii') species coexistence 
$(a_r,b_r)$ with $a_r = (1 - \mu / \rho \lambda) \, \sigma / \lambda$ and 
$b_r = \mu / \lambda$, which comes into existence and is linearly stable if the
predation rate exceeds the threshold $\lambda_c$.
The finite carrying capacity causes the Jacobian matrix eigenvalues to acquire 
negative real parts, indicating an exponential approach to the stable fixed 
point $(a_r,b_r)$:
\begin{equation}
  \epsilon_\pm = - \frac{\mu \, \sigma}{2 \, \rho \, \lambda} \left[ 1 \pm 
  \sqrt{1 - \frac{4 \, \rho \, \lambda}{\sigma} \left( 
  \frac{\rho \, \lambda}{\mu} - 1 \right)} \right] \ .
\label{lvrese}
\end{equation}
The neutral cycles of the unrestricted model (\ref{lvreqa}) are thus replaced
either by a stable node for which $\epsilon_\pm$ are both real, namely when 
$\sigma > \sigma_s = 4 \lambda \, \rho \, (\rho \lambda / \mu - 1) > 0$, or 
alternatively $\mu / \rho < \lambda < \lambda_s = (1 + \sqrt{1 + \sigma / \mu})
\, \mu / 2 \rho$; or by a stable focus with complex conjugate stability matrix 
eigenvalue pair, and consequent spiraling relaxation towards the fixed point if
$\sigma < \sigma_s$ or $\lambda > \lambda_s$.
In this situation, both predator and prey populations approach their stationary
values $(a_r,b_r)$ via damped oscillations.
Adding spatial degrees of freedom, finite local carrying capacities can be
implemented in a lattice model through limiting the maximum occupation number 
per site for each species.
Most drastically, one can permit at most a single particle per lattice site 
(as, for example, implemented in Ref.~\cite{Mobilia07}); the binary predation 
reaction then has to occur between predators and prey on nearest-neighbor
sites, and new offspring is to be placed on adjacent positions.
In that case, one may actually dispense with hopping processes, since all 
particle production reactions automatically entail population spreading.
Upon adding diffusive spreading terms (with diffusivities $D_A$, $D_B$) to the 
mean-field rate equations, one may describe spreading activity fronts of prey 
invading empty regions followed by predators feeding on them.
A well-established lower bound for the front propagation speed is 
\cite{Dunbar83, Sherratt97, Murray02}
\begin{equation}
  v_{\rm front} > \sqrt{4 \, D_A \, (\lambda \, \rho - \mu)} \ .
\label{frtspd}
\end{equation}
 
To summarize, within the mean-field rate approximation, a finite prey carrying 
capacity $\rho$, which can be viewed as the mean result of local restrictions 
on the prey density originating from limited resources, crucially changes the 
phase diagram:
There emerges an extinction threshold (at $\lambda_c$ for fixed $\mu$) for the 
predator population, which in a spatially extended system in the thermodynamic
and infinite-time limit becomes a genuine continuous nonequilibrium phase 
transition from an active to an absorbing state. 
In addition, deep in the species coexistence phase the restricted 
Lotka--Volterra model is characterized by transient decaying population 
oscillations, which become overdamped upon approaching the predator extinction
threshold.

\subsection{Monte Carlo simulation results in the species coexistence phase}

Various authors have studied individual-based stochastic lattice predator-prey 
models, predominantly in two dimensions and typically with periodic boundary 
conditions, that in the well-mixed mean-field limit reduce to the classical 
Lotka--Volterra system; see Refs.~\cite{Matsuda92}--\cite{Kowalik02} and 
\cite{Washenberger07} for a partial listing. 
The following is a concise summary of some fundamental results from these
extensive numerical investigations, as pertinent for the subsequent 
field-theoretic analysis.

Monte Carlo simulations in two dimensions, both in the absence and presence of 
local density limitations, observe the emergence of prominent spatio-temporal 
structures associated with remarkably strong fluctuations in the species
coexistence phase, even far away from the continuous nonequilibrium predator 
extinction transition.
Spherically expanding growth fronts of prey closely followed by predators 
periodically sweep the system; any small surviving clusters of prey then serve 
as nucleation centers for new population waves that subsequently interact and 
for large population densities eventually merge with each other.
These spreading activity fronts are especially sharp for the site-restricted
model variants, whereas in simulation runs performed with arbitrarily many 
particles per site, the fronts appear more diffuse and localized \cite{Movies}.
Equal-time density correlation functions can be employed to determine the 
spatial width $\sim 10 \ldots 20$ lattice sites of the spreading activity 
regions \cite{Mobilia07, Washenberger07, Dobramysl08}.
In comparison with the mean-field bound (\ref{frtspd}), the front velocity was 
measured to be typically enhanced by a factor up to $\sim 2 \ldots 3$ in 
simulation runs starting with a single localized activity seed 
\cite{Dobramysl08}.

Averaging over the weakly coupled and periodically emerging structures yields
long-lived but damped population oscillations.
As the system size increases, one observes the relative oscillation amplitudes 
to decrease; in the thermodynamic limit, the quasi-periodic population
fluctuations eventually terminate.
Yet locally density oscillations persist for both predators and prey species.
In the absence of spatial degrees of freedom, these can be understood by 
performing a van Kampen system size expansion about the absorbing steady 
state \cite{McKane05}.
The fluctuation corrections may then essentially be described by means of a 
damped harmonic oscillator driven by white noise that will on occasion 
resonantly incite large-amplitude excursions away from the stable fixed point 
in the phase plane.

From the prominent peaks detected in the Fourier-transformed concentration 
signals, characteristic oscillation frequencies can be inferred 
\cite{Mobilia07, Washenberger07, Dobramysl08}.
The thus numerically determined typical population oscillation frequencies are
found to be reduced by a factor ranging between $2$ and $6$ (depending on the
other rates) in the stochastic spatially extended system as compared to the
mean-field prediction, a considerable downward renormalization obviously caused
by fluctuations and reaction-induced spatio-temporal correlations; compare 
Fig.~9 in Ref.~\cite{Mobilia07} and Fig.~6(b) in Ref.~\cite{Washenberger07}.
However, the measured oscillation frequencies $f$ roughly follow the 
square-root dependence on the rates $\mu$ and $\sigma$ suggested by the 
linearized mean-field approximation: $\omega_0 = 2\pi f_0 = \sqrt{\mu \sigma}$,
yet with noticeable deviations once either $\sigma$ or $\mu$ significantly 
differ from unity.
In addition, the functional dependence of $f$ on the rates $\mu$ and $\sigma$
is surprisingly similar, at least in a mid-range interval of values for both
rates near $1$.
As we shall see in section~\ref{flcorr}, these observations and quantitative 
trends are remarkably accurately reproduced by a first-order analytic 
perturbation theory for fluctuation corrections in stochastic lattice 
Lotka--Volterra models.

As the predation efficiency $\lambda$ is reduced (with all other parameters 
held constant), stochastic lattice Lotka--Volterra systems with site occupation
restrictions display just the same qualitative scenarios as revealed by the 
mean-field analysis for eqs.~(\ref{lvreqc}) with finite prey carrying capacity:
First, the focal stationary points in the phase plane are replaced by stable
nodes (corresponding to real stability matrix eigenvalues); the population 
oscillations then cease, and no interesting spatial structures are discernible
aside from hardly fluctuating localized clusters of predators in a `sea' of
prey that almost fill the entire lattice \cite{Movies}.
At a sufficiently small critical value $\lambda_c$, at last the predator 
extinction threshold is encountered, and the measured critical scaling laws 
near this active- to absorbing state transition are very well described by the 
accepted critical exponents of directed percolation 
\cite{Tome94}--\cite{Kowalik02}. 

Simulations in one dimension (usually on a circular domain) yield a crucial 
difference between model variants that incorporate or neglect site occupation 
number restrictions: 
In the former situation, the $A$ and $B$ particles quickly segregate into
distinct domains, with the predation reactions occurring only at their 
boundaries.
The long-time evolution is consequently dictated by the very slow coarsening of
merging predator domains \cite{Mobilia07}.
In contrast, in the absence of site occupation restrictions, one observes the 
system to invariably remain in an active fluctuating coexistence state 
\cite{Washenberger07}.

We finally remark that the above statements naturally all pertain to 
sufficiently large lattices. 
Of course, any finite system with an absorbing steady state will in principle
eventually reach and remain trapped in it. 
However, the associated tpyical extinction times are understood to grow fast 
with system size, namely according to a power law 
\cite{Parker09, Dobrinevski12}; simulation runs performed in reasonably large 
lattices consequently never reach this extinction state during their entire 
duration.

\section{Field-theoretic analysis}

This section will first provide a brief overview how a coherent-state path 
integral representation can be constructed directly from the fundamental master
equation that defines a stochastic interacting particle system 
\cite{Doi76}--\cite{Tauber05}, see also Refs.~\cite{Tauber07, Tauber09};
Ref.~\cite{Andreanov06} provides an illustrative alternative derivation.
This field-theoretic representation faithfully encodes statistical fluctuations, 
including those caused by discreteness and the internal reaction noise, as well 
as emerging correlations in spatial reaction-diffusion systems, and allows for 
systematic approximative analysis, as will be detailed below for two-species 
predator-prey models.
For the sake of completeness, the essential steps of mapping spatial stochastic
Lotka--Volterra models with site occupation number restrictions near the 
predator extinction threshold onto Reggeon field theory \cite{Mobilia07} will 
be repeated here as well.
The subsequent section~\ref{flcorr} is then concerned with fluctuation 
corrections in the two-species coexistence phase, which become manifest 
through propagator renormalizations.
 
\subsection{Field theory representation}

The Doi--Peliti approach is based on the fact that at any time the 
configurations in locally reacting particle systems can be enumerated through 
specifying the occupation numbers of each species per lattice site $i$, say 
here $n_i$ for the predators $A$ and $m_i$ for the prey $B$, and that the 
effect of any allowed stochastic process is to merely modify these on-site 
integer occupation numbers.
For now, arbitrarily many particles of either species are allowed to occupy any
lattice point: $n_i, m_i = 0,1,\ldots,\infty$.
The master equation for our local reaction scheme (\ref{lvreac}) that governs 
the time evolution of the configurational probability to find $n_i$ predators 
and $m_i$ prey on site $i$ at time $t$ through the balance of gain and loss 
terms reads
\begin{eqnarray} 
  &&\frac{\partial}{\partial t} \, P(n_i,m_i;t) = \mu 
  \Bigl[ (n_i+1) \, P(n_i+1,m_i;t) -  n_i \, P(n_i,m_i;t) \Bigr] \nonumber \\
  &&\qquad\qquad\qquad 
  + \sigma \Bigl[ (m_i-1) \, P(n_i,m_i-1;t) - m_i \, P(n_i,m_i;t) \Bigr]
\label{masloc} \\
  &&\qquad + \lambda' \Bigl[ (n_i - 1) \, (m_i + 1) \, P(n_i-1,m_i+1;t) 
  - n_i \, m_i \, P(n_i,m_;t) \Bigr] \ . \nonumber
\end{eqnarray} 
As initial condition, we may for instance choose a Poisson distribution
$P(n_i,m_i;0) = \overline{n}_0^{n_i} \, \overline{m}_0^{m_i} \, 
e^{- \overline{n}_0 - \overline{m}_0} / n_i! \, m_i!$ with mean initial 
predator and prey concentrations $\overline{n}_0$ and $\overline{m}_0$.

Because all reactions just change the site occupation numbers by integer 
values, a Fock space representation is particularly useful.
To this end, we introduce the bosonic ladder operator algebra 
$[ a_i , a_j ] = 0$, $[ a_i , a_j^\dagger ] = \delta_{ij}$ for species $A$, 
from which we construct the predator particle number eigenstates 
$| n_i \rangle$, $a_i \, |n_i \rangle = n_i \, |n_i-1 \rangle$, 
$a_i^\dagger \, |n_i \rangle = |n_i + 1 \rangle$,
$a_i^\dagger \, a_i \, |n_i \rangle = n_i \, |n_i \rangle$.
A Fock state with $n_i$ particles on site $i$ is obtained from the empty
`vacuum' configuration $| 0 \rangle$, defined via $a_i \, | 0 \rangle = 0$, 
through $| n_i \rangle =  {a_i^\dagger}^{n_i} | 0 \rangle$.
In the same manner, we proceed for the prey particles, with associated 
annihilation and creation operators $b_i$ and $b_i^\dagger$ that all commute 
with the predator ladder operators: $[a_i , b_j ] = 0 = [a_i , b_j^\dagger ]$.

To implement the stochastic kinetics for the entire lattice, one considers the
master equation for the configurational probability $P(\{ n_i\},\{ m_i \};t)$,
given by a sum over all lattice points of the right-hand side of 
eq.~(\ref{masloc}), and recognizes that a general Fock state is constructed by 
the tensor product 
$| \{ n_i \}, \{ m_i \} \rangle = \prod_i |n_i \rangle \, |m_i \rangle$.
One then defines a time-dependent formal state vector through a linear
combination of all possible Fock states, weighted by their configurational
probability at time $t$:
\begin{equation}
  | \Phi(t) \rangle = \sum_{\{ n_i \}, \{ m_i \}} P(\{ n_i \},\{ m_i \};t) \,
  | \{ n_i \}, \{ m_i \} \rangle \ .
\label{stavec}
\end{equation}
This superposition state thus encodes the stochastic temporal evolution.
Straightforward manipulations now transform the time dependence from the linear
master equation into an `imaginary-time' Schr\"odinger equation, governed by a 
time-independent stochastic Liouville time evolution operator $H$:
\begin{equation}   
  \frac{\partial | \Phi(t) \rangle}{\partial t} = - H \, | \Phi(t) \rangle \ , 
  \quad {\rm or} \quad | \Phi(t) \rangle = e^{- H\, t} \, | \Phi(0) \rangle \ .
\label{itschr}
\end{equation}
For on-site reactions, 
$H_{\rm reac} = \sum_i H_r(a_i^\dagger, b_i^\dagger;a_i,b_i)$ is a sum of local
(normal-ordered) contributions; for the Lotka--Volterra predator-prey system 
one obtains
\begin{equation}  
  H_{\rm reac} = - \sum_i \left[ \mu \, \bigl( 1 - a_i^\dagger \bigr) \, a_i 
  + \sigma \, \bigl( b_i^\dagger - 1 \bigr) \, b_i^\dagger\,  b_i + \lambda' \,
  \bigl( a_i^\dagger - b_i^\dagger \bigr) \, a_i^\dagger \, a_i \, b_i \right] 
  \ .
\label{lovham}
\end{equation}
Note that each reaction process is represented by two contributions, 
originating respectively from the gain and loss terms in the master equation.
For nearest-neighbor hopping of particles $A (B)$ with rate $D'_A (D'_B)$ 
between neighboring lattice sites $\langle i j \rangle$, one finds the 
additional contributions
\begin{equation}  
  H_{\rm diff} = \sum_{<ij>} \left[ D'_A \, \bigl( a_i^\dagger - a_j^\dagger
  \bigr) \, \bigl( a_i - a_j \bigr)  + D_B' \, \bigl( b_i^\dagger - b_j^\dagger
  \bigr) \, \bigl( b_i - b_j \bigr) \right] \ .
\label{difham}
\end{equation} 

Our goal is to compute averages and correlation functions with respect to the 
configurational probability $P(\{ n_i \}, \{ m_i \};t)$\, which is accomplished
by means of the projection state 
$\langle {\cal P} | = \langle 0 | \prod_i e^{a_i} e^{b_i}$, for which 
$\langle {\cal P} | 0 \rangle = 1$ and $\langle {\cal P} | a_i^\dagger = 
\langle {\cal P} | = \langle {\cal P} | b_i^\dagger$, since 
$[ e^{a_i} , a_j^\dagger ] = e^{a_i} \, \delta_{ij}$.
For the desired statistical averages of observables ${\cal O}$, which naturally
must all be expressible as functions of the occupation numbers $n_i$ and $m_i$,
one obtains 
\begin{equation}
  \fl \quad \langle {\cal O}(t) \rangle = \! \sum_{\{ n_i \}, \{ m_i \}} \!
  {\cal O}(\{ n_i \}, \{ m_i \}) \, P(\{ n_i \}, \{ m_i \};t) = \langle 
  {\cal P} | \, {\cal O}(\{ a_i^\dagger \, a_i \},\{ b_i^\dagger \, b_i\}) \, 
  | \Phi(t) \rangle \ .
\label{masave}
\end{equation}
As a consequence of probability conservation, one finds for ${\cal O} = 1$:
$1 = \langle {\cal P} | \Phi(t) \rangle = \langle {\cal P} | e^{- H \, t} | 
\Phi(0) \rangle$.
Therefore $\langle {\cal P} | H = 0$ must hold; upon commuting 
$e^{\sum_i (a_i + b_i)}$ with $H$, the creation operators are effectively 
shifted by $1$: 
$a_i^\dagger \to 1 + a_i^\dagger$, $b_i^\dagger \to 1 + b_i^\dagger$. 
The probability conservation condition is thus satisfied provided
$H_i(a_i^\dagger \to 1,b_i^\dagger \to 1;a_i,b_i) = 0$, which is of course true
for our explicit expressions (\ref{lovham}) and (\ref{difham}).
Through this prescription, we may replace $a_i^\dagger \, a_i \to a_i$ and
$b_i^\dagger \, b_i \to b_i$  in all averages; e.g., the predator and prey 
densities become $a(t) = \langle a_i(t) \rangle$ and 
$b(t) = \langle b_i(t) \rangle$.

In the bosonic operator representation above, we have assumed that no 
restrictions apply to the particle occupation numbers $n_i$ on each site.
If $n_i \leq 2 s + 1$, one may instead employ a representation in terms of spin
$s$ operators.
An alternative approach, devised by van~Wijland, utilizes the bosonic theory, 
but incorporates site occupation restrictions through explicit constraints, 
which ultimately appear as exponentials in the number operators 
\cite{Wijland01}.
For example, limiting the local prey occupation numbers to $0$ or $1$ modifies
the birth process in (\ref{lovham}) to $H_{i \, \sigma} = \sigma \, 
\bigl( 1 - b_i^\dagger \bigr) \, b_i^\dagger\,  b_i \, e^{- b_i^\dagger b_i}$.
Instead, one could also just add a reaction that restricts the local population
numbers, e.g., $B + B \to B$ with rate $\nu'$, yielding an additional term
$H_{i \, \nu'} = - \nu' \, \bigl( 1 - b_i^\dagger \bigr) \, b_i^\dagger \, 
b_i^2$.

As a next step, we follow a well-established route in quantum many-particle 
theory \cite{Negele88} and proceed towards a field theory representation via
constructing the path integral equivalent to the `Schr\"odinger' dynamics 
(\ref{itschr}) based on coherent states, which are defined as right eigenstates
of the annihilation operators, 
$a_i \, | \alpha_i \rangle = \alpha_i \, | \alpha_i \rangle$ and 
$b_i \, | \beta_i \rangle = \beta_i \, | \beta_i \rangle$, labeled by their 
complex eigenvalues $\alpha_i$ and $\beta_i$.
One readily confirms the explicit formula $|\alpha_i \rangle = 
\exp(- \frac12 \, |\alpha_i|^2 + \alpha_i \, a_i^\dagger) | 0 \rangle$, the  
overlap integral $\langle \alpha_j | \alpha_i \rangle = 
\exp(- \frac12 |\alpha_i|^2 - \frac12 |\alpha_j|^2 + \alpha_j^* \, \alpha_i )$,
and the (over-)completeness relation $\int \prod_i d^2 \alpha_i \, 
|\{ \alpha_i \} \rangle \, \langle \{ \alpha_i \}| = \pi$.
Splitting the temporal evolution (\ref{itschr}) into infinitesimal increments,
inserting the (over-)completeness relation at each time step, and further
straightforward manipulations (details can be found in Ref.~\cite{Tauber05}) 
eventually yield an expression for the configurational average 
\begin{equation} 
  \langle {\cal O}(t) \rangle \! \propto \! \int \! \prod_i \! d\alpha_i \, 
  d\alpha_i^* \, d\beta_i \, d\beta_i^* \, {\cal O}(\{ \alpha_i \} , \{ 
  \beta_i \}) \, \exp (- S[\alpha_i^*, \beta_i^*; \alpha_i, \beta_i;t])  \, , \
\label{copain}
\end{equation}
with an exponential statistical weight that is determined by the `action'
\begin{eqnarray}
  &&S[\alpha_i^*,\beta_i^*; \alpha_i,\beta_i; t] = \sum_i \Biggl[ \,\int_0^t \!
  dt' \, \Biggl( \alpha_i^* \, \frac{\partial \alpha_i}{\partial t'} 
  + \beta_i^* \, \frac{\partial \beta_i}{\partial t'} 
  + H_r(\alpha_i^*,\beta_i^*;\alpha_i,\beta_i) \Biggr) \nonumber \\
  &&\qquad\qquad\qquad\qquad\qquad\quad - \alpha_i(t) - \beta_i(t) 
  - \overline{n}_0 \, \alpha^*_i(0) - \overline{m}_0 \, \beta^*_i(0) \, \Biggr]
  \ ,
\label{cohact} 
\end{eqnarray}
where the second term at the final time $t$ stems from the projection states, 
while the last one originates in the initial Poisson distributions.
Through this procedure, in the original quasi-Hamiltonian the creation and 
annihilation operators $a_i^\dagger (b_i^\dagger)$ and $a_i (b_i)$ are at each
time instant replaced with the complex numbers $\alpha_i^* (\beta_i^*)$ and 
$\alpha_i (\beta_i)$.

Finally, we proceed to take the continuum limit, 
$\sum_i \to a_0^{-d} \int d^dx$, $\alpha_i(t) \to a_0^d \, a(\vec{x},t)$, 
$\beta_i(t) \to a_0^d \, b(\vec{x},t)$, where $a_0$ denotes the original 
microscopic lattice constant, whereupon the continuous fields $a$ and $b$ 
acquire dimensions of particle densities, and 
$\alpha_i^*(t) \to {\hat a}(\vec{x},t)$, 
$\beta_i^*(t) \to {\hat b}(\vec{x},t)$, such that ${\hat a}$ and ${\hat b}$
remain dimensionless.
The `bulk' part of the action then becomes
\begin{equation} 
  \fl \quad S[{\hat a},{\hat b}; a,b] = \int \! d^dx \! \int \! dt \, \Biggl[ 
  {\hat a} \, \Biggl( \frac{\partial}{\partial t} - D_A \nabla^2 \! \Biggr) a 
  + {\hat b} \, \Biggl( \frac{\partial}{\partial t} - D_B \nabla^2 \! \Biggr) b
  + H_r({\hat a},{\hat b}; a,b) \Biggr] \, ,  
\label{masfth} 
\end{equation}
where the discrete hopping contribution (\ref{difham}) has turned into a 
continuum diffusion term, with diffusivities $D_{A/B} = a_0^2 \, D'_{A/B}$.
We have thus arrived at a (mesoscopic) field theory for stochastic
reaction-diffusion processes, with its dynamics governed by  two independent 
fields for each particle species, without invoking any assumptions on the form 
of the internal reaction noise.
For the Lotka--Volterra reactions (\ref{lvreac}) with site occupation number
restrictions and/or population-limiting reactions with diffusive spreading in 
$d$ spatial dimensions, the bulk action (\ref{masfth}) reads explicitly 
\cite{Mobilia07}
\begin{eqnarray}
  &&S[{\hat a},{\hat b}; a,b] = \int \! d^dx \int \! dt \, \Biggl[ {\hat a} \, 
  \Biggl( \frac{\partial}{\partial t} - D_A \nabla^2 \Biggr) a + {\hat b} \, 
  \Biggl( \frac{\partial}{\partial t} - D_B \nabla^2 \Biggr) b 
\label{rlvact} \\
  &&\qquad\quad + \mu \bigl( {\hat a} - 1 \bigr) \, a - \sigma \bigl( 
  {\hat b} - 1 \bigr) \, {\hat b} \, b  \, e^{- a_0^d \, {\hat b} \, b}
  + \nu \bigl( {\hat b} - 1 \bigr) \, {\hat b} \, b^2 - \lambda \bigl( 
  {\hat a} - {\hat b} \bigr) \, {\hat a} \, a \, b \, \Biggr] , \nonumber
\end{eqnarray}
with $\nu = a_0^d \, \nu'$ and $\lambda = a_0^d \, \lambda'$; for unrestricted 
site occupation numbers, the exponential term just needs to be replaced with 
$1$, and $\nu$ set to $0$.
Expanding $e^{- a_0^d \, {\hat b} \, b} \approx 1 - a_0^d \, {\hat b} \, b$ in 
the limit $a_0 \to 0$ effectively replaces the `hard' exponential constraint 
with a `softened' particle number restriction, which will henceforth be used. 
The action (\ref{rlvact}) may now serve as a basis for further systematic 
coarse-graining, constructing a perturbation expansion as described below, or,
if required, a subsequent renormalization group analysis 
\cite{Tauber05, Tauber07, Tauber09}.

The associated classical field equations follow from the stationarity 
conditions $\delta S / \delta a = 0 = \delta S / \delta b$, which are always 
solved by ${\hat a} = 1 = {\hat b}$, reflecting probability conservation, and 
$\delta S / \delta {\hat a}({\vec x},t) = 0 
= \delta S / \delta {\hat b}({\vec x},t)$, which yields precisely the 
mean-field rate equations augmented by diffusion terms.
Setting ${\hat a} = 1 = {\hat b}$, for $a_0 = 0 = \nu$ one indeed arrives at 
the Lotka--Volterra rate equations (\ref{lvreqa}), without any restrictions on 
the prey population density, plus diffusive spreading.
The modified prey density equation (\ref{lvreqc}) with diffusion follows 
instead, if either $\nu = 0$ and a `soft' particle number restriction is 
implemented with the natural identification $\rho = a_0^{-d}$, or alternatively
with $a_0 = 0$ but adding a pair coagulation reaction with rate 
$\nu = \sigma / \rho$.
It is thus convenient to perform the field shift 
${\hat a}({\vec x},t) = 1 + {\tilde a}({\vec x},t)$, 
${\hat b}({\vec x},t) = 1 + {\tilde b}({\vec x},t)$, whereupon the action 
becomes, 
\begin{eqnarray}
  &&S[{\tilde a},{\tilde b};a,b] = \int \! d^dx \int \! dt \, \Biggl[ 
  {\tilde a} \, \Biggl( \frac{\partial}{\partial t} - D_A \, \nabla^2 
  + \mu \Biggr) a + {\tilde b} \, \Biggl( \frac{\partial}{\partial t} 
  - D_B \, \nabla^2 - \sigma \Biggr) b \nonumber \\
  &&\quad\qquad\qquad\qquad\quad - \sigma \, {\tilde b}^2 \, b + 
  \frac{\sigma}{\rho} \, (1 + {\tilde b})^\alpha \, {\tilde b} \, b^2 - \lambda
  \, (1 + {\tilde a}) \, ({\tilde a} - {\tilde b}) \, a \, b \, \Biggr] ,
\label{slvact} 
\end{eqnarray}
with integer $\alpha = 2$ parametrizing a softened restricted prey occupation,
whereas $\alpha = 1$ captures instead the presence of the binary reaction 
$B + B \to B$; the unrestricted model is of course recovered for 
$\rho \to \infty$.

We remark that for $\alpha = 1$, the action (\ref{slvact}) is equivalent to the
two coupled Langevin stochastic equations of motion
\begin{eqnarray}
  &&\frac{\partial a({\vec x},t)}{\partial t} =  (D_A \, \nabla^2 - \mu) \,
  a({\vec x},t) + \lambda \, a({\vec x},t) \, b({\vec x},t) + \zeta({\vec x},t)
  \ ,
\label{langeq} \\
  &&\frac{\partial b({\vec x},t)}{\partial t} = (D_B \, \nabla^2  + \sigma) \,
  b({\vec x},t) - \frac{\sigma}{\rho} \, b({\vec x},t)^2 
  - \lambda \, a({\vec x},t) \, b({\vec x},t) + \eta({\vec x},t) \ , \nonumber 
\end{eqnarray}
i.e., the diffusive rate equations for the local particle densities, with added
Gaussian white noise with vanishing means, 
$\langle \zeta \rangle = 0 = \langle \eta \rangle$, and the 
(cross-)correlations
\begin{eqnarray}
  &&\langle \zeta({\vec x},t) \, \zeta({\vec x}',t') \rangle = 2 \lambda \, 
  a({\vec x},t) \, b({\vec x},t) \, \delta({\vec x}-{\vec x}') \, \delta(t-t') 
  \ , \nonumber \\
  &&\langle \zeta({\vec x},t) \, \eta({\vec x}',t') \rangle = - \lambda \, 
  a({\vec x},t) \, b({\vec x},t) \, \delta({\vec x}-{\vec x}') \, \delta(t-t')
  \ ,
\label{noicor} \\
  &&\langle \eta({\vec x},t) \, \eta({\vec x}',t') \rangle = 2 \sigma \, 
  b({\vec x},t) \, \Bigl[1 - b({\vec x},t) / \rho \Bigr] \, 
  \delta({\vec x}-{\vec x}') \, \delta(t-t') \ , \nonumber
\end{eqnarray}
describing multiplicative noise terms that vanish with the particle densities, 
as appropriate for the absorbing state at $a = 0 = b$.
Similar Langevin equations were derived in Ref.~\cite{Butler09}.

The equivalence of eqs.~(\ref{langeq}) and (\ref{noicor}) with the action
(\ref{slvact}) follows immediately from the standard Janssen--De~Dominicis 
field theory representation of Langevin dynamics 
\cite{Janssen76, DeDominicis76} (see also Refs.~\cite{Tauber07, Tauber09}),
according to which the set of Langevin equations 
$\partial s_i({\vec x},t) / \partial t = F_i[\{ s_i({\vec x},t) \}] +
\zeta_i({\vec x},t)$ with $\langle \zeta_i \rangle = 0$ and noise correlations 
$\langle \zeta_i({\vec x},t) \, \zeta_j({\vec x}',t') \rangle = 
2 L_{i j}[\{ s_i({\vec x},t) \}] \, \delta({\vec x}-{\vec x}') \, \delta(t-t')$
is governed by the action
\begin{equation}
  \fl \qquad\quad S[\{ {\tilde s}_i \}; \{ s_i \}] = \int \! d^dx \int \! dt \,
  \sum_i \, \Biggl[ {\tilde s}_i \, \Biggl( \frac{\partial s_i}{\partial t} \, 
  - F_i[\{ s_i \}] \Biggr) - \sum_j {\tilde s}_i \, L_{i j}[\{ s_i \}] \, 
  {\tilde s}_j \Biggr] \ .
\label{langac}
\end{equation}
For $\alpha = 2$, the action (\ref{slvact}) contains a cubic term of the 
`auxiliary' field ${\tilde b}$, and a direct Langevin representation is not 
obviously possible.
In the following, the field theory action (\ref{slvact}) will serve as the 
starting point for further manipulations (i) to briefly recapitulate the 
identification of critical directed percolation as the universality class that 
governs the continuous active to absorbing state transition at the predator 
extinction threshold \cite{Mobilia07}, and (ii) to compute the 
fluctuation-induced renormalization to lowest order in the predation rate for
the population oscillation frequency and damping, as well as the diffusion 
coefficient in the two-species coexistence phase \cite{Tauber11}.

\subsection{Predator extinction transition and Reggeon field theory}

Here we provide the basic steps by which the effective field theory that
describes the universal scaling behavior near the predator extinction threshold
is constructed, following Ref.~\cite{Mobilia07}.
For $\lambda \approx \lambda_c = \mu / \rho$, very few predators remain, 
while the prey almost fill the entire lattice, $a({\vec x},t) \approx a_r = 0$,
$b({\vec x},t) \approx b_r = \rho$.
The reaction scheme (\ref{lvreac}) is thus essentially replaced with
$A \to \emptyset$ and $A \to A + A$.
We then also need to add a growth-limiting process for the predator population,
for example again through the binary coagulation reaction $A + A \to A$, say 
with rate $\tau$; heuristically, we have then already arrived at the standard
single-species death-birth-annihilation reactions that in essence define 
directed percolation processes (see, e.g., 
Refs.~\cite{Tauber05, Janssen05, Tauber07, Tauber09}).

In the Doi--Peliti representation, we consequently transform the action 
(\ref{slvact}) to new fluctuating prey fields 
$e({\vec x},t) = \rho - b({\vec x},t)$ with vanishing mean 
$\langle e \rangle = 0$, and 
${\tilde e}({\vec x},t) = - {\tilde b}({\vec x},t)$. 
With the additional predator pair coagulation reaction, this yields
\begin{eqnarray}
  &&S[{\tilde a},{\tilde e};a,e] = \int \! d^dx \int \! dt \, \Biggl[ 
  {\tilde a} \, \Biggl( \frac{\partial}{\partial t} - D_A \, \nabla^2 + \mu 
  - \lambda \, \rho \Biggr) a + \tau \, {\tilde a} \, (1 + {\tilde a}) \, a^2  
  \nonumber \\
  &&\quad + {\tilde e} \, \Biggl( \frac{\partial}{\partial t} - D_B \nabla^2 
  + \sigma \Biggr) e - \sigma \, \Bigl[ (1 - {\tilde e})^\alpha - 1 \Bigr] 
  {\tilde e} \, (\rho - 2 e) - \frac{\sigma}{\rho} \, (1 - {\tilde e})^\alpha 
  \, {\tilde e} \, e^2 \nonumber \\
  &&\qquad\qquad\qquad\qquad - \lambda \, \rho \Bigl( {\tilde a}^2 
  + (1 + {\tilde a}) \, {\tilde e} \Bigr) \, a + \lambda \, (1 + {\tilde a}) \,
  ({\tilde a} + {\tilde e}) \, a \, e \, \Biggr] \, . 
\label{dpact1}
\end{eqnarray}
Next we note that the birth rate is a relevant parameter in the renormalization
group sense, which scales to infinity under scale transformations; this
observation simply expresses the fact that fluctuations of the nearly uniform 
prey population become strongly suppressed through the `mass' term 
$\propto \sigma$ for the $e$ fields.
It is therefore appropriate to introduce rescaled fields 
$\phi({\vec x},t) = \sqrt{\sigma} \, e({\vec x},t)$ and 
${\tilde \phi}({\vec x},t) = \sqrt{\sigma} \, {\tilde e}({\vec x},t)$, and 
subsequently take the limit $\sigma \to \infty$, which yields the much reduced 
effective action
\begin{eqnarray}
  &&S_\infty[{\tilde a},{\tilde \phi};a,\phi] = \int \! d^dx \int \! dt \, 
  \Biggl[ {\tilde a} \, \Biggl( \frac{\partial}{\partial t} - D_A \, \nabla^2 
  + \mu - \lambda \, \rho \Biggr) a  - \lambda \, \rho \, {\tilde a}^2 \, a 
  \nonumber \\
  &&\qquad\qquad\qquad\qquad\qquad\qquad + \tau \, {\tilde a} \, 
  (1 + {\tilde a}) \, a^2 + {\tilde \phi} \, \phi 
  + \alpha \, \rho \, {\tilde \phi}^2 \, \Biggr] \, .
\label{dpact2}
\end{eqnarray}
Since the fields $\phi$ and ${\tilde \phi}$ only appear as a bilinear form in
the action (\ref{dpact2}), they can readily be integrated out, leaving 
\begin{equation}
  \fl \quad S'_\infty[{\widetilde {\psi}},\psi] = \int \! d^dx \int \! dt \, 
  \Biggl[ {\widetilde \psi} \, \Biggl( \frac{\partial}{\partial t} 
  + D_A \, \Bigl( r_A - \nabla^2 \Bigr) \! \Biggr) \, \psi 
  - u \, {\widetilde \psi} \, \Bigl( {\widetilde \psi} - \psi \Bigr) \, \psi 
  + \tau \, {\widetilde \psi }^2 \, \psi^2 \Biggr] ,
\label{dprft}
\end{equation}
where $\psi({\vec x},t) = a({\vec x},t) \, \sqrt{\tau / \lambda \, \rho}$, 
${\widetilde \psi}({\vec x},t) = {\tilde a}({\vec x},t) \, 
\sqrt{\lambda \, \rho / \tau}$, $r_A = (\mu - \lambda\, \rho) / D_A$, and 
$u = \sqrt{\tau \, \lambda \, \rho}$.
This new effective nonlinear coupling $u$ becomes dimensionless at $d_c = 4$, 
signifying the upper critical dimension for this field theory. 
Near four dimensions, the quartic term $\propto \tau$ constitutes an irrelevant
contribution in the renormalization group sense and may be omitted for the
analysis of universal asymptotic power laws at the phase transition.
The action (\ref{dprft}) then becomes identical to Reggeon field theory, which
is known to describe the critical scaling exponents for directed percolation
\cite{Obukhov80}--\cite{Janssen05}. 
This mapping to Reggeon field theory hence confirms the general expectation
that the predator extinction threshold is governed by the directed percolation 
universality class \cite{Tome94, Boccara94}, \cite{Albano99}--\cite{Monetti00},
\cite{Antal01, Kowalik02}, 
which features quite prominently in phase transitions to absorbing states 
\cite{Janssen81, Grassberger82}, even in multi-species systems 
\cite{Janssen01}. 
The universal scaling properties of critical directed percolation are 
well-understood and quantitatively characterized to remarkable accuracy, both 
numerically through extensive Monte Carlo simulations and analytically by means
of renormalization group calculations (for overviews, see 
Refs.~\cite{Hinrichsen00, Odor04, Janssen05}).

\section{Fluctuation corrections in the coexistence phase}
\label{flcorr}

We now proceed to investigate and analyze the effect of intrinsic stochastic
fluctuations and spatial correlations in the two-species coexistence phase (and 
in the thermodynamic limit), by means of a systematic perturbation expansion 
about the (undamped) mean-field theory with infinite prey carrying capacity, 
$\rho \to \infty$.
Various additional technical details are deferred to the three Appendices.

\subsection{Doi--Peliti action in the two-species coexistence phase}

In order to address fluctuation corrections in the predator-prey coexistence
phase \cite{Tauber11}, we start again from the Doi--Peliti field theory action 
(\ref{slvact}), and introduce proper fluctuating fields 
$c({\vec x},t) = a({\vec x},t) - \langle a \rangle$ and 
$d({\vec x},t) = b({\vec x},t) - \langle b \rangle$ with vanishing mean:
\begin{equation}
  \fl \qquad\qquad a({\vec x},t) = \frac{\sigma}{\lambda} \biggl( 1 - 
  \frac{\mu}{\rho \, \lambda} + A_c \biggr) + c({\vec x},t) \ , \quad
  b({\vec x},t) =  \frac{\mu}{\lambda} \, (1 + B_c) + d({\vec x},t) \ .
\label{cphcd}
\end{equation}
Here, the mean-field values for the stationary densities have been taken into
account already, such that the counter-terms $A_c$ and $B_c$, which are 
naturally determined by the conditions 
$\langle c \rangle = 0 = \langle d \rangle$, contain only fluctuation 
contributions; this is in accord with standard procedures for perturbation
expansions in ordered phases \cite{Brezin73}--\cite{Tauber92}.
Upon inserting (\ref{cphcd}) into (\ref{slvact}), and renaming 
${\tilde a}({\vec x},t) = {\tilde c}({\vec x},t)$ and 
${\tilde b}({\vec x},t) = {\tilde d}({\vec x},t)$, one arrives at the action 
$S[{\tilde c},{\tilde d};c,d]$ in terms of the new fields.
It is a sum of three contributions,
\begin{eqnarray}
  &&S_s[{\tilde c},{\tilde d};c,d] = - \, \frac{\sigma \, \mu}{\lambda} \int \!
  d^dx \int \! dt \, \Biggl[ \, B_c \biggl( 1 - \frac{\mu}{\rho \, \lambda} + 
  A_c \biggr) \, {\tilde c} \nonumber \\
  &&\quad - (1 + B_c) \, \biggl( A_c + \frac{\mu}{\rho \, \lambda} \, B_c 
  \biggr) \, {\tilde d} + \biggl( 1 - \frac{\mu}{\rho \, \lambda} + A_c \biggr)
  \, (1 + B_c) \, {\tilde c} \, ({\tilde c} - {\tilde d}) \nonumber \\
  &&\quad + (1 + B_c) \, \biggl[ 1 - \alpha \, \frac{\mu}{\rho \, \lambda} \, 
  (1 + B_c) \biggr] \, {\tilde d}^2 - (\alpha - 1) \, 
  \frac{\mu}{\rho \, \lambda} \, (1 + B_c)^2 \, {\tilde d}^3 \, \Biggr] \, ,
\label{sffsr} 
\end{eqnarray}
which represent source terms, the bilinear or harmonic contributions
\begin{eqnarray}
  &&\!\!\!\! S_h[{\tilde c},{\tilde d};c,d] = \int \! d^dx \! \int \!\! dt \, 
  \Biggl[ \, {\tilde c} \, \Biggl( \frac{\partial}{\partial t} - D_A \, 
  \nabla^2 - \mu \, B_c \Biggr) \, c + \mu \, (1 + B_c) \, {\tilde d} \, c 
\label{sffbl} \\
  &&- \sigma \, \biggl( 1 - \frac{\mu}{\rho \, \lambda} + A_c \biggr) \, 
  {\tilde c} \, d + {\tilde d} \, \Biggl( \frac{\partial}{\partial t} - D_B \,
  \nabla^2 + \sigma \, \biggl[ A_c + \frac{\mu}{\rho \, \lambda} \, (1 + 2 B_c)
  \biggr] \Biggr) \, d \, \Biggr] \, , \nonumber
\end{eqnarray}
and finally the nonlinear vertices
\begin{eqnarray}
  &&S_v[{\tilde c},{\tilde d};c,d] = - \int \! d^dx \int \! dt \, \Biggl[ \,
  \mu \, (1 + B_c) \, {\tilde c} \, ({\tilde c} - {\tilde d}) \, c \nonumber \\
  &&\quad + \sigma \, \biggl( 1 - \frac{\mu}{\rho \, \lambda} + A_c \biggr) \, 
  {\tilde c} \, ({\tilde c} - {\tilde d}) \, d + \sigma \, \biggl[ 1 - 2 \alpha
  \, \frac{\mu}{\rho \, \lambda} \, (1 + B_c) \biggr] \, {\tilde d}^2 \, d 
\label{sffnl} \\
  &&\quad - 2 (\alpha - 1) \, \frac{\sigma \, \mu}{\rho \, \lambda} \, 
  (1 + B_c) \, {\tilde d}^3 \, d + \lambda \, (1 + {\tilde c}) \, 
  ({\tilde c} - {\tilde d}) \, c \, d - \frac{\sigma}{\rho} \, 
  (1 + {\tilde d})^\alpha \, {\tilde d} \, d^2 \, \Biggr] \nonumber 
\end{eqnarray}
(recall that the exclusion parameter assumes only the values $\alpha = 1$ or 
$2$).
Note that since the definitions (\ref{cphcd}) already contain the mean-field 
expectation values of the field, the linear source terms 
$\sim {\tilde c}, {\tilde d}$ in (\ref{sffsr}) are mere counter-terms.

The integrand in the harmonic action (\ref{sffbl}) can be written as a bilinear
form $({\tilde c} \ {\tilde d}) \, {\overline A} \, {c \choose d}$.
Defining the spatial and temporal Fourier transform for an arbitrary field via
\begin{equation}
  \phi({\vec x},t) = \int \frac{d^dq}{(2 \pi)^d} \int \frac{d \omega}{2 \pi}
  \ \phi({\vec q},\omega) \, e^{i ({\vec q} \cdot {\vec x} - \omega t)} \ ,
\label{ftdef}
\end{equation}
(and omitting the fluctuation corrections $\sim A_c, B_c$), we have in Fourier
space
\begin{equation}
  {\overline A}(q,\omega) = \left( \begin{array}{cc} - i \omega + D_A \, q^2 &
  - \sigma \, (1 - \mu / \rho \, \lambda) \\ \mu & - i \omega + D_B \, q^2 
  + \sigma \, \mu / \rho \, \lambda \end{array} \right) \ .
\label{harcl}
\end{equation}
The next step is to diagonalize the non-symmetric bilinear coupling matrix
${\bar A} = {\overline A}(0,0)$.
To this end, we need its right and left eigenvectors, 
${\bar A} \, e_\pm = {\bar \lambda}_\pm \, e_\pm$, 
$f_\pm^T \, {\bar A} = {\bar \lambda}_\pm \, f_\pm^T$ that satisfy 
the orthogonality relation $f_\pm^T \, e_\mp = 0$.
Introducing the eigenvector matrices $P = (e_+ \ e_-)$ and $Q = (f_+ \ f_-)$, 
one then readily confirms 
$Q^T {\bar A} \, P = {\rm diag}(\lambda_+ \ \lambda_-)$, with the diagonal 
elements $\lambda_\pm = {\bar \lambda}_\pm \, f_\pm^T \, e_\pm$.
Upon defining new fields $\varphi_\pm$ and ${\widetilde \varphi}_\pm$ via 
${c \choose d} = P \, {\varphi_+ \choose \varphi_-}$ and $({\tilde c} \ 
{\tilde d}) = ({\widetilde \varphi}_+ \ {\widetilde \varphi}_-) \, Q^T$, 
finally $({\tilde c} \ {\tilde d}) \, {\bar A} \, {c \choose d} = 
({\widetilde \varphi}_+ \ {\widetilde \varphi}_-) \, {\rm diag}(\lambda_+ , 
\lambda_-) \, {\varphi_+ \choose \varphi_-} = \lambda_+ {\widetilde \varphi}_+ 
\varphi_+ +  \lambda_- {\widetilde \varphi}_- \varphi_-$.
The eigenvalues of the matrix ${\bar A}$ are just the negative of the stability
matrix eigenvalues in the coexistence phase, 
${\bar \lambda}_\pm = \pm i \omega_0 + \gamma_0 = - \epsilon_\pm$, c.f.
eq.~(\ref{lvrese}), with the mean-field (`bare') oscillation frequency 
$\omega_0$ and damping constant $\gamma_0$ (see also Ref.~\cite{Butler09}):
\begin{equation}
   \omega_0^2 = \mu \, \sigma \left( 1 - \frac{\mu}{\rho \, \lambda} \right) 
   - \gamma_0^2 \ , \quad 
   \gamma_0 = \frac{\mu \, \sigma}{2 \, \rho \, \lambda} \ .
\label{bfrdm}
\end{equation}
Observe that $\omega_0^2 = \mu \, \sigma$ and $\gamma_0 \to 0$ as the carrying
capacity $\rho \to \infty$: 
There is no damping of the mean-field oscillations in the absence of local
particle number restrictions; in this situation, damping terms are in fact 
generated by stochastic fluctuations, as will be demonstrated below.
Choosing the eigenvectors 
$e_\pm^T = (i \omega_0 \mp \gamma_0 \ \pm \mu) / i \omega_0 \sqrt{2 \mu}$,
$f_\pm^T = (\pm \mu \ i \omega_0 \pm \gamma_0) / i \omega_0 \sqrt{2 \mu}$, 
with $f_\pm^T \, e_\pm = \pm 1 / i \omega_0$, the harmonic action (\ref{sffbl})
is diagonalized by means of the linear field transformations
\begin{eqnarray}
   &&c = \frac{1}{\sqrt{2 \mu}} \, \biggl( \, \varphi_+ + \varphi_-
   - \gamma_0 \, \frac{\varphi_+ - \varphi_-}{i \omega_0} \biggr) \, , \quad 
   d = \sqrt{\frac{\mu}{2}} \ \frac{\varphi_+ - \varphi_-}{i \omega_0} 
   \nonumber \\ 
   &&{\tilde c} = \sqrt{\frac{\mu}{2}} \ \frac{{\widetilde \varphi}_+ 
   - {\widetilde \varphi}_-}{i \omega_0} \, , \quad 
   {\tilde c} = \frac{1}{\sqrt{2 \mu}} \, \biggl( \, {\widetilde \varphi}_+ + 
   {\widetilde \varphi}_- + \gamma_0 \, 
   \frac{{\widetilde \varphi}_+ - {\widetilde \varphi}_-}{i \omega_0} \biggr)
   \, .
\label{cphph}
\end{eqnarray}

Indeed, upon inserting (\ref{cphph}) into (\ref{sffbl}), one obtains the 
harmonic action in terms of the new fields 
\begin{eqnarray}
  &&\!\!\!\!\!\!\!\!\!\!\!\!\!\! S_h[{\widetilde \varphi}_\pm;\varphi_\pm]
  = \frac{1}{i \omega_0} \int \! d^dx \! \int \!\! dt \, \Biggl[ \, 
  {\widetilde \varphi}_+ \Biggl( \frac{\partial}{\partial t} - D_0 \, \nabla^2 
  + \frac{\gamma_0}{i \omega_0} \, D_0' \, \nabla^2 + i \omega_0 + \gamma_0 
  \nonumber \\
  &&+ \frac{i \omega_0 + \gamma_0 - \mu}{2 i \omega_0} \, \sigma A_c
  + \frac{(i \omega_0 - \gamma_0)(i \omega_0 + \gamma_0 - \mu) + 4 \gamma_0 
  (i \omega_0 + \gamma_0)}{2 i \omega_0} \, B_c \Biggr) \, \varphi_+ \nonumber 
  \\
  &&\!\!\!\!\!\!\!\!\!\!\!\!\!\! - {\widetilde \varphi}_+ \Biggl( 
  \frac{i \omega_0 + \gamma_0}{i \omega_0} \, D_0' \nabla^2 + 
  \frac{i \omega_0 + \gamma_0 - \mu}{2 i \omega_0} \, \sigma A_c
  - \frac{(i \omega_0 + \gamma_0)(i \omega_0 - 3 \gamma_0 - \mu)}{2 i \omega_0}
  \, B_c \Biggr) \, \varphi_- \nonumber \\
  &&\!\!\!\!\!\!\!\!\!\!\!\!\!\! + {\widetilde \varphi}_- \Biggl( 
  \frac{i \omega_0 - \gamma_0}{i \omega_0} \, D_0' \nabla^2 + 
  \frac{i \omega_0 - \gamma_0 + \mu}{2 i \omega_0} \, \sigma A_c
  + \frac{(i \omega_0 - \gamma_0)(i \omega_0 + 3 \gamma_0 + \mu)}{2 i \omega_0}
  \, B_c \Biggr) \, \varphi_+ \nonumber \\
  &&- {\widetilde \varphi}_- \Biggl( \frac{\partial}{\partial t} - D_0 \, 
  \nabla^2 - \frac{\gamma_0}{i \omega_0} \, D_0' \, \nabla^2 - i \omega_0 
  + \gamma_0 + \frac{i \omega_0 - \gamma_0 + \mu}{2 i \omega_0} \, \sigma A_c
\label{sphibl} \\
  &&\qquad\qquad\qquad\qquad - \frac{(i \omega_0 + \gamma_0)(i \omega_0 
  - \gamma_0 + \mu) - 4 \gamma_0 (i \omega_0 - \gamma_0)}{2 i \omega_0} \, B_c 
  \Biggr) \, \varphi_- \Biggr] \, , \nonumber
\end{eqnarray}
where $D_0 = (D_A + D_B) / 2$ denotes the mean particle diffusivity, and
$D_0' = (D_A - D_B) / 2$ indicates the asymmetry in the diffusion coefficients.
In the following, we shall restrict ourselves to the case of equal 
diffusivities $D_A = D_B = D_0$ and $D_0' = 0$; the harmonic propagators in the
diagonalized theory then read in Fourier space
\begin{equation}
\fl \qquad \langle {\widetilde \varphi}_\pm({\vec q},\omega) \, 
  \varphi_\pm({\vec q}',\omega') \rangle_0 = 
  \frac{\pm i \omega_0}{- i \omega + D_0 \, q^2 \pm i \omega_0 + \gamma_0} \ 
  (2 \pi)^{d+1} \, \delta({\vec q} + {\vec q}') \, \delta(\omega + \omega') \ ,
\label{props}
\end{equation}
whereas the off-diagonal two-point correlation functions 
$\langle {\widetilde \varphi}_\pm({\vec q},\omega) \, 
\varphi_\mp({\vec q}',\omega') \rangle$ contain only counter-terms and hence
vanish in the harmonic approximation.
Akin to spin waves in magnets, the poles of the propagators (\ref{props}) 
describe (anti-)clockwise propagating waves with frequency $\omega_0$ and 
damping $\gamma_0$, with additional diffusive relaxation $\sim D_0 \, q^2$.
The delta functions in (\ref{props}) reflect spatial and temporal time 
translation invariance.

Upon expressing the sources (\ref{sffsr}) and nonlinear contributions 
(\ref{sffnl}) as functionals of the new fields, a multitude of terms is
generated which renders any subsequent analysis quite cumbersome, see 
\ref{fulfth}.
Consequently we shall address the limit of large prey carrying capacity 
$\rho \to \infty$, for which the mean-field approximation predicts undamped
oscillatory modes with frequency $\omega_0 = \sqrt{\mu \sigma}$, see 
eq.~(\ref{lvrese}).
Correspondingly, we shall henceforth retain finite $\rho$ and non-zero damping 
$\gamma_0$ solely in the propagator terms (\ref{sphibl}), but set 
$\rho^{-1} = 0 = \gamma_0$ everywhere else.
The source terms then just read
\begin{eqnarray}
  &&S_s[{\widetilde \varphi}_\pm;\varphi_\pm] = \int \! d^dx \! \int \!\! dt \,
  \Biggl[ \, \sqrt{\frac{\sigma}{2}} \, \frac{1}{i \lambda} \, \Biggl( \biggl[ 
  i \omega_0 \, A_c \, (1 + B_c) - \mu \, B_c \, (1 + A_c) \biggr] \, 
  {\widetilde \varphi}_+ \nonumber \\
  &&\qquad\qquad\qquad\qquad\qquad\qquad\quad 
  + \biggl[ i \omega_0 \, A_c \, (1 + B_c) \, + \mu \, B_c \, (1 + A_c) \biggr]
  \, {\widetilde \varphi}_- \! \Biggr) \nonumber \\
  &&\qquad\qquad\qquad\qquad\quad\ - \frac{1 + B_c}{2 \, \lambda} \, \Biggl( 
  \biggl[ (i \omega_0 - \mu) \, (1 + A_c) + \sigma \biggr] \, 
  {\widetilde \varphi}_+^2
\label{sphirr} \\
  &&\qquad\qquad\ + 2 \, \biggl[ \mu \, (1 + A_c) + \sigma \biggr] \, 
  {\widetilde \varphi}_+ \, {\widetilde \varphi}_- - \biggl[ (i \omega_0 + \mu)
  \, (1 + A_c) - \sigma \biggr] \, {\widetilde \varphi}_-^2 \Biggr) \Biggr] \, 
  , \nonumber
\end{eqnarray}
compare (\ref{sphisr}) in \ref{fulfth}.
The linear source terms are mere counter-terms; following eq.~(\ref{langac}),
one may interpret the quadratic ones as generated by stochastic noise.

From eq.~(\ref{sffnl}) one obtains the three-point vertices in the limit 
$\rho \to \infty$:
\begin{eqnarray}
  &&S_v[{\widetilde \varphi}_\pm;\varphi_\pm] = 
  \frac{1}{2 \sqrt{2 \mu} \, \omega_0^2} \, \int \! d^dx \! \int \!\! dt \, 
  \Biggl[ \Biggl( (i \omega_0 - \mu) \Bigl[ i \omega_0 \, (1 + A_c) - \mu \, 
  (1 + B_c) \Bigr] \nonumber \\
  &&\qquad\qquad\qquad\qquad\qquad\qquad\qquad\qquad\qquad\qquad
  \qquad\quad + i \omega_0 \, \sigma \Biggr) \, {\widetilde \varphi}_+^2 \, 
  \varphi_+ \nonumber \\
  &&\qquad\qquad\qquad\ - \Biggl( (i \omega_0 - \mu) \Bigl[ i \omega_0 \, 
  (1 + A_c) + \mu \, (1 + B_c) \Bigr] + i \omega_0 \, \sigma \Biggr) \, 
  {\widetilde \varphi}_+^2 \, \varphi_- \nonumber \\
  &&\qquad\qquad\qquad\quad\ + 2 \, \Biggl( \mu \Bigl[ i \omega_0 \, (1 + A_c) 
  - \mu \, (1 + B_c) \Bigr] + i \omega_0 \, \sigma \Biggr) \, 
  {\widetilde \varphi}_+ \, {\widetilde \varphi}_- \, \varphi_+ \nonumber \\  
  &&\qquad\qquad\qquad\quad\ - 2 \, \Biggl( \mu \Bigl[ i \omega_0 \, (1 + A_c) 
  + \mu \, (1 + B_c) \Bigr] + i \omega_0 \, \sigma \Biggr) \, 
  {\widetilde \varphi}_+ \, {\widetilde \varphi}_- \, \varphi_- \nonumber \\
  &&\qquad\qquad\qquad\ - \Biggl( (i \omega_0 + \mu) \Bigl[ i \omega_0  \, 
  (1 + A_c) - \mu \, (1 + B_c) \Bigr] - i \omega_0 \, \sigma \Biggr) \, 
  {\widetilde \varphi}_-^2 \, \varphi_+ \nonumber \\
  &&\qquad\qquad\qquad\ + \Biggl( (i \omega_0 + \mu) \Bigl[ i \omega_0 \, 
  (1 + A_c) + \mu \, (1 + B_c) \Bigr] - i \omega_0 \, \sigma \Biggr) \, 
  {\widetilde \varphi}_-^2 \, \varphi_- \nonumber \\
  &&\qquad\qquad - \lambda \, (i \omega_0 - \mu) \, {\widetilde \varphi}_+ \, 
  (\varphi_+^2 - \varphi_-^2) - \lambda \, (i \omega_0 + \mu) \, 
  {\widetilde \varphi}_- \, (\varphi_+^2 - \varphi_-^2) \Biggr] \ . 
\label{nlvrr3}
\end{eqnarray}
The nonlinear vertices of the full action are listed in eqs.~(\ref{nlver3}), 
(\ref{vert45}) in \ref{fulfth}.
Note that in the large carrying capacity approximation, the various field 
theory contributions naturally become independent of the parameter $\alpha$.
Both the reduced and full actions remain essentially invariant under exchange
of the labels $+ \longleftrightarrow -$, aside from complex conjugation, an 
obvious consequence of the complex conjugate eigenvalue pairs 
${\bar \lambda}_\pm$ for ${\overline A}$ and the corresponding eigenvector 
symmetry, see eq.~(\ref{cphph}).
Formally, this symmetry is conveniently expressed in terms the vertex functions
with $m_\pm$ external outgoing  ${\widetilde \varphi_\pm}$ and $n_\pm$ 
incoming $\varphi_\pm$ legs:
\begin{equation}
  \Gamma_{+^{m_+} \, -^{m_-} ; \, +^{n_+} \, -^{n_-}}(\vec{x}_i,t_i) =
  \Gamma_{+^{m_-} \, -^{m_+} ; \, +^{n_-} \, -^{n_+}}(\vec{x}_i,t_i)^* \ .
\label{vfsym}
\end{equation}
As a direct consequence, $\Gamma_{+^m \, -^m ; \, +^n \, -^n}(\vec{x}_i,t_i)$ 
must be real.

\subsection{Counter-terms and propagator renormalization to one-loop order}

The propagators (\ref{props}) along with two two-point noise sources 
(\ref{sphirr}) and three-point vertices (\ref{nlvrr3}) represent the building 
blocks for the Feynman diagrams that graphically represent the different 
contributions in a perturbation expansion with respect to the predation rate 
$\lambda$, which serves as the nonlinear coupling here \cite{Tauber11}.

\begin{figure}[ht]
\begin{center}
\includegraphics[width=9cm]{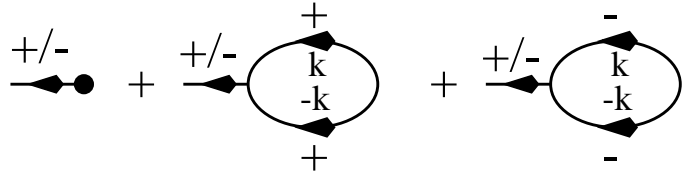}
\end{center}
\caption{Feynman graphs for the expectation values 
  $\langle \varphi_\pm \rangle$ up to one-loop order, where the `$\bullet$' 
  symbol in the first diagram represents the counter-terms.}
\label{lvfig1}
\end{figure}
As our first step, we need to determine the counter-terms $A_c$ and $B_c$ to 
first order in $\lambda$ from the conditions $\langle \varphi_\pm \rangle = 0$.
The associated Feynman graphs are shown in Fig.~\ref{lvfig1}, and the 
corresponding analytic expressions read
\begin{eqnarray}
  &&0 = \sqrt{\frac{\sigma}{2}} \, \frac{i}{\lambda} \, 
  \biggl[ i \omega_0 \, A_c \mp \mu \, B_c \biggr] \nonumber \\
  &&\qquad + \frac{i \omega_0 \mp \mu}{4 \sqrt{2 \mu}} \int \! 
  \frac{d^dk}{(2\pi)^d} 
  \left( \frac{\mu - \sigma - i \omega_0}{i \omega_0 + \gamma_0 + D_0 k^2} - 
  \frac{\mu - \sigma + i \omega_0}{- i \omega_0 + \gamma_0 + D_0 k^2} \right) .
\label{exp1lp}
\end{eqnarray}
These are readily solved, with the result
\begin{eqnarray}
  &&A_c = B_c = \frac{i \lambda}{4 \omega_0} \int\!\! \frac{d^dk}{(2\pi)^d} 
  \left( \frac{\mu - \sigma - i \omega_0}{i \omega_0 + \gamma_0 + D_0 k^2} - 
  \frac{\mu - \sigma + i \omega_0}{- i \omega_0 + \gamma_0 + D_0 k^2} \right)
  \!+ O(\lambda^2) \nonumber \\
  &&\qquad\quad\,\ = \frac{\lambda}{2} \int \! \frac{d^dk}{(2\pi)^d} \ 
  \frac{\mu - \sigma + \gamma_0 + D_0 k^2}
  {\omega_0^2 + (\gamma_0 + D_0 k^2)^2} + O(\lambda^2) \, .
\label{cnt1lp}
\end{eqnarray}

We may now proceed to the fluctuation renormalization of the propagators 
(\ref{props}) to first order in the predation rate $\lambda$.
To this end, we require the two-point vertex functions 
$\Gamma_{\pm ; \pm}({\vec q},\omega)$ to one-loop order.
Denoting their fluctuation corrections by 
$\Gamma_{\pm ; \pm}^{(1)}({\vec q},\omega)$, the structure of the low-frequency
and small-wavevector expansion is
\begin{eqnarray}
  &&\Gamma_{\pm ; \pm}({\vec q},\omega) = 1 
  + {\rm Re} \, \Gamma_{\pm ; \pm}^{(1)}(0,0) \pm \frac{\gamma_0}{i \omega_0} 
  + i \, {\rm Im} \, \Gamma_{\pm ; \pm}^{(1)}(0,0) 
  \pm \frac{D_0 \, q^2}{i \omega_0} \pm \frac{\omega}{\omega_0} \nonumber \\
  &&\qquad\qquad\qquad + q^2 \, 
  \frac{\partial \, \Gamma_{\pm ; \pm}^{(1)}({\vec q},0)}{\partial \, q^2} 
  \Bigg\vert_{{\vec q} = 0} + i \omega \, 
  \frac{\partial \, \Gamma_{\pm ; \pm}^{(1)}(0,\omega)}{\partial \, i \omega} 
  \Bigg\vert_{\omega = 0} + \ldots
\label{g11exp}
\end{eqnarray}
Note that the symmetry (\ref{vfsym}) implies
\begin{equation}
  \Gamma_{+ ; +}({\vec q},0) = \Gamma_{- ; -}({\vec q},0)^* \, , \quad
  \frac{\partial \, \Gamma_{+ ; +}({\vec q},\omega)}{\partial \, i \omega} 
  \Bigg\vert_{\omega = 0} = \left( 
  \frac{\partial \, \Gamma_{- ; -}({\vec q},\omega)}{\partial \, i \omega} 
  \Bigg\vert_{\omega = 0} \, \right)^* \! .
\label{g11id}
\end{equation}

\begin{figure}[ht]
\begin{center}
\includegraphics[width=15.5cm]{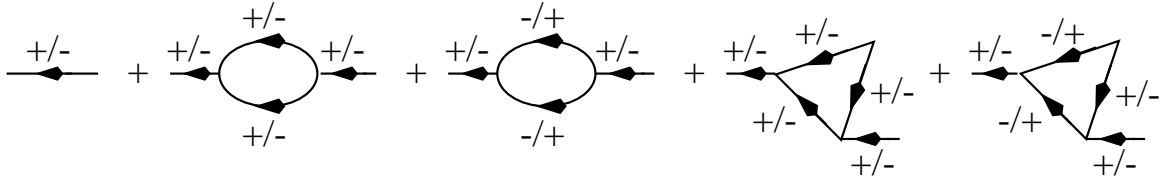}
\end{center}
\caption{Feynman graphs for the vertex functions 
  $\Gamma_{\pm ; \pm}({\vec q},\omega)$ up to one-loop order.}
\label{lvfig2}
\end{figure}
The one-loop Feynman diagrams for $\Gamma_{\pm ; \pm}({\vec q},\omega)$ are 
depicted in Fig.~\ref{lvfig2}.
Performing the internal frequency integrals, one arrives at the associated 
analytic expressions
\begin{eqnarray}
  &&\!\!\!\!\!\!\!\!\!\!\!\!\!\! \Gamma_{\pm ; \pm}({\vec q},\omega) = 
  \frac{1}{\pm i \omega_0} \, \Bigl[ i \omega \pm i \omega_0 + \gamma_0 + D_0 
  \, q^2 + \Bigl( \frac{\sigma - \mu}{2} \pm i \omega_0 \Bigr) \, A_c \Bigr] 
  \nonumber \\
  &&\!\!\!\!\!\!\!\!\!\! 
  + \frac{\lambda \, (\pm i \omega_0 - \mu)}{8 \, \mu \, \omega_0^2} \, [\mu 
  \, (\sigma - \mu \pm 2 i \omega_0) \mp i \omega_0 \, \sigma] \! \int \!\!
  \frac{(2 \pi)^{-d} \, d^dk}{i \omega / 2 \pm i \omega_0 + \gamma_0 + D_0 \, 
  (q^2 / 4 + k^2)} \nonumber \\
  &&\quad + \frac{\lambda \, (\pm i \omega_0 - \mu)}{8 \, \mu \, \omega_0^2} \,
  [\mu \, (\sigma + \mu) \pm i \omega_0 \, \sigma] \int \! \frac{(2 \pi)^{-d} 
  \, d^dk}{i \omega / 2 \mp i \omega_0 + \gamma_0 + D_0 \, (q^2 / 4 + k^2)}
  \nonumber \\
  &&\pm \frac{\lambda}{8 \, i \omega_0} \, (\sigma - \mu \pm 2 i \omega_0) \,
  (\sigma - \mu \pm i \omega_0) \int \! \frac{d^dk}{(2 \pi)^{d}} \
  \frac{1}{\pm i \omega_0 + \gamma_0 + D_0 \, ({\vec q} / 2 + {\vec k})^2} 
  \nonumber \\
  &&\qquad\qquad \times 
  \frac{1}{\pm i \omega_0 + \gamma_0 + D_0 \, ({\vec q} / 2 - {\vec k})^2} \
  \frac{\pm i \omega_0 + \gamma_0 + D_0 \, (q^2 / 4 + k^2)}
  {i \omega / 2 \pm i \omega_0 + \gamma_0 + D_0 (q^2 / 4 + k^2)} \nonumber \\
  &&\quad \mp \frac{\lambda}{8 \, i \omega_0} \, (\sigma + \mu)^2 \int \! 
  \frac{d^dk}{(2 \pi)^{d}} \ 
  \frac{1}{\gamma_0 + D_0 \, ({\vec q} / 2 + {\vec k})^2} \
  \frac{1}{\gamma_0 + D_0 \, ({\vec q} / 2 - {\vec k})^2} \nonumber \\
  &&\qquad\qquad\qquad\qquad\qquad\qquad \times 
  \frac{\gamma_0 + D_0 \, (q^2 / 4 + k^2)}
  {i \omega / 2 \mp i \omega_0 + \gamma_0 + D_0 (q^2 / 4 + k^2)} \ , 
\label{prp1lp}
\end{eqnarray}
where the last two terms have been symmetrized with respect to the external
wavevector ${\vec q}$.
Naturally, eq.~(\ref{prp1lp}) satisfies the symmetry constraints (\ref{g11id}).
Clearly, ${\rm Im} \, \Gamma_{\pm ; \pm}^{(1)}(0,0)$ does not vanish, which
implies that the nonlinear fluctuations generically either generate a damping 
term for the population oscillations, see eq.~(\ref{rendam}), or induce an
instability towards spatial structure formation, as observed in the lattice Monte 
Carlo simulations.

Notice furthermore the convolution of both clock- and anti-clockwise 
propagating modes in the `triangular' fluctuation loop of the last Feynman 
graph in Fig.~\ref{lvfig2}.
As a consequence, the imaginary `mass' terms $\pm i \omega_0$ in the first two
factors within the associated wavevector integral cancel each other, as becomes
apparent in the final term of eq.~(\ref{prp1lp}).
For vanishing damping $\gamma_0 \to 0$ this induces an infrared divergence in 
$d \leq 2$ dimensions.
It is precisely these contributions that cause large fluctuation corrections 
for the renormalized oscillation frequency, eq.~(\ref{intfre}) below, in the 
coexistence phase of the spatial Lotka--Volterra system even at finite (but 
small) damping $\gamma_0$.

\subsection{Renormalized damping, oscillation frequency, and diffusion 
  coefficient}

Appropriate definitions of renormalized oscillation parameters are suggested by
the functional form (\ref{g11exp}) of the vertex functions 
$\Gamma_{\pm ; \pm}({\vec q},\omega)$.
We thus cast the renormalized two-point vertex functions in the form
\begin{equation}
  \Gamma^R_{\pm ; \pm}({\vec q},\omega) = 1 \pm \frac{\gamma_R}{i \omega_R}
  \pm \frac{\omega}{\omega_R} \pm \frac{D_R \, q^2}{i \omega_R} \ ,
\label{renprp}
\end{equation}
whence we identify the renormalized damping $\gamma_R$, frequency $\omega_R$, 
and diffusivity $D_R$ via
\begin{eqnarray}
  &&\gamma_R = 
  \frac{\gamma_0 \mp \omega_0 \, {\rm Im} \, \Gamma_{\pm ; \pm}^{(1)}(0,0)}
  {1 \mp \omega_0 \, {\rm Im}\, [\partial \, \Gamma_{\pm ; \pm}^{(1)}(0,\omega)
  / \partial \, i \omega]_{\omega = 0}} \ , 
\label{rendam} \\
  &&\omega_R = \frac{\omega_0\, [1 + {\rm Re}\, \Gamma_{\pm ; \pm}^{(1)}(0,0)]}
  {1 \mp \omega_0 \, {\rm Im}\, [\partial \, \Gamma_{\pm ; \pm}^{(1)}(0,\omega)
  / \partial \, i \omega]_{\omega = 0}} \ ,
\label{renfre} \\
  &&D_R = \frac{D_0 \mp \omega_0 \, {\rm Im} \, [\partial \, 
  \Gamma_{\pm ; \pm}^{(1)}({\vec q},0) / \partial \, q^2]_{{\vec q} = 0}}
  {1 \mp \omega_0 \, {\rm Im}\, [\partial \, \Gamma_{\pm ; \pm}^{(1)}(0,\omega)
  / \partial i \omega]_{\omega = 0}} \ .
\label{rendif}
\end{eqnarray}
Note that a negative `damping' $\gamma_R < 0$ in eq.~(\ref{renprp}) indicates 
an instability towards a spatially inhomogeneous configuration at wavenumber 
$q_c = \sqrt{|\gamma_R| / D_R}$ or characteristic wavelength 
$\lambda_c = 2 \pi \sqrt{D_0 / |\gamma_R|} + O(\lambda^2)$.

Upon evaluating the basic one-loop result (\ref{prp1lp}), and following the 
prescriptions (\ref{rendam})--(\ref{rendif}), it is a straightforward task to
compute the renormalized parameters $\gamma_R$, $\omega_R$, and $D_R$.
Intermediate steps and technical details can be found in \ref{gameva}.
As a final task, one needs to perform the resulting wavevector integrals for the 
fluctuation corrections.
With the aid of the integral table in \ref{inttab}, one finds with (\ref{kint1}), 
(\ref{dint1}), and (\ref{dint3}) for the renormalized or fluctuation-induced 
damping (\ref{1lpdam}):
\begin{eqnarray}
  &&\gamma_R = \gamma_0 + \lambda \, \frac{\Gamma(1-d/2)}{2^{d+3} \, \pi^{d/2}}
  \, \Bigl( \frac{\omega_0}{D_0} \Bigr)^{d/2} \,
  \Biggl( \frac{\sigma}{\mu} + \frac{\mu}{\sigma} \Biggr) \, {\rm Im} \, 
  \Bigl( \frac{\gamma_0}{\omega_0} + i \Bigr)^{-1+d/2} \nonumber \\
  &&\qquad\quad\ + \lambda \, \frac{\Gamma(2-d/2)}{2^{d+3} \, \pi^{d/2}} \, 
  \Bigl( \frac{\omega_0}{D_0} \Bigr)^{d/2} \,
  \Biggl[ \Biggl( \frac{\sigma}{\mu} + \frac{\mu}{\sigma} - 4 \Biggr) \, 
  {\rm Re} \, \Bigl( \frac{\gamma_0}{\omega_0} + i \Bigr)^{-2+d/2} \nonumber \\
  &&\qquad\qquad\qquad\qquad\quad\quad\ 
  - 3 \, \Biggl( \sqrt{\frac{\sigma}{\mu}} - \sqrt{\frac{\mu}{\sigma}} \, 
  \Biggr) \, {\rm Im} \, \Bigl( \frac{\gamma_0}{\omega_0} + i \Bigr)^{-2+d/2} 
  \Biggr] + O(\lambda^2) \nonumber \\
  &&\quad\ = \gamma_0 + \lambda \, \Bigl( \frac{\omega_0}{D_0} \Bigr)^{d/2} \,
  \Delta {\tilde \gamma}_R + O(\lambda^2) \, . 
\label{intdam}
\end{eqnarray}
The renormalized oscillation frequency (\ref{1lpfre}) becomes
\begin{eqnarray}
  &&\omega_R = \omega_0 + \lambda \, \frac{\Gamma(1-d/2)}{2^{d+4} \, \pi^{d/2}}
  \, \Bigl( \frac{\omega_0}{D_0} \Bigr)^{d/2} \, \Biggl[
  \Biggl( \frac{\sigma}{\mu} + \frac{\mu}{\sigma} + 2 \Biggr) \, 
  {\rm Re} \, \Bigl( \frac{\gamma_0}{\omega_0} + i \Bigr)^{-1+d/2} \nonumber \\
  &&\qquad\qquad\qquad + 4 \, \sqrt{\frac{\sigma}{\mu}} \ {\rm Im} \, 
  \Bigl( \frac{\gamma_0}{\omega_0} + i \Bigr)^{-1+d/2} - 
  \Biggl( \frac{\sigma}{\mu} + \frac{\mu}{\sigma} + 2 \Biggr) 
  \Bigl( \frac{\gamma_0}{\omega_0} \Bigr)^{-1+d/2} \Biggr] \nonumber \\  
  &&\qquad\quad\;\ + \lambda \, \frac{\Gamma(2-d/2)}{2^{d+4} \, \pi^{d/2}} \, 
  \Bigl( \frac{\omega_0}{D_0} \Bigr)^{d/2} \, \Biggl[ 4 \, 
  \Biggl( \sqrt{\frac{\sigma}{\mu}} - \sqrt{\frac{\mu}{\sigma}} \, \Biggr) \,
  {\rm Re} \, \Bigl( \frac{\gamma_0}{\omega_0} + i \Bigr)^{-2+d/2} \nonumber \\
  &&\qquad\qquad\qquad\qquad\qquad\qquad\qquad
  + \Biggl( \frac{\sigma}{\mu} + \frac{\mu}{\sigma} - 4 \Biggr) \, 
  {\rm Im} \, \Bigl( \frac{\gamma_0}{\omega_0} + i \Bigr)^{-2+d/2} \Biggr] 
  \nonumber \\
  &&\qquad\quad\;\ + \lambda \, \frac{\Gamma(3-d/2)}{2^{d+5} \, \pi^{d/2}} \, 
  \Bigl( \frac{\omega_0}{D_0} \Bigr)^{d/2} \,
  \Biggl[ \Biggl( \frac{\sigma}{\mu} + \frac{\mu}{\sigma} - 4 \Biggr) \, 
  {\rm Re} \, \Bigl( \frac{\gamma_0}{\omega_0} + i \Bigr)^{-3+d/2} \nonumber \\
  &&\qquad\qquad\qquad\qquad\qquad 
  - 3 \, \Biggl( \sqrt{\frac{\sigma}{\mu}} - \sqrt{\frac{\mu}{\sigma}} \, 
  \Biggr) \, {\rm Im} \, \Bigl( \frac{\gamma_0}{\omega_0} + i \Bigr)^{-3+d/2} 
  \Biggr] + O(\lambda^2) \nonumber \\
  &&\quad\ = \omega_0 + \lambda \, \Bigl( \frac{\omega_0}{D_0} \Bigr)^{d/2} \,
  \Delta {\tilde \omega}_R + O(\lambda^2) \, ,
\label{intfre}
\end{eqnarray}
while the renormalized diffusion coefficient (\ref{1lpdif}) reads
\begin{eqnarray}
  &&D_R = D_0 + \lambda \, \frac{\Gamma(1-d/2)}{d \, 2^{d+3} \, \pi^{d/2}}
  \, \Bigl( \frac{\omega_0}{D_0} \Bigr)^{-1+d/2} 
  \Biggl( \frac{\sigma}{\mu} + \frac{\mu}{\sigma} + 2 \Biggr) \, 
  {\rm Im} \, \Bigl( \frac{\gamma_0}{\omega_0} + i \Bigr)^{d/2} \nonumber \\
  &&\qquad\qquad\! - \lambda \, \frac{\Gamma(1-d/2)}{2^{d+4} \, \pi^{d/2}}
  \, \Bigl( \frac{\omega_0}{D_0} \Bigr)^{-1+d/2} 
  \Biggl( \frac{\sigma}{\mu} + \frac{\mu}{\sigma} + 2 \Biggr) \,
  {\rm Re} \, \Bigl( \frac{\gamma_0}{\omega_0} + i \Bigr)^{-1+d/2} \nonumber \\
  &&\qquad\quad\, - \lambda \, \frac{\Gamma(2-d/2)}{2^{d+5} \, \pi^{d/2}} \, 
  \Bigl( \frac{\omega_0}{D_0} \Bigr)^{-1+d/2} \, \Biggl[ 2 \, 
  \Biggl( \sqrt{\frac{\sigma}{\mu}} - \sqrt{\frac{\mu}{\sigma}} \, \Biggr) \,
  {\rm Re} \, \Bigl( \frac{\gamma_0}{\omega_0} + i \Bigr)^{-2+d/2} \nonumber \\
  &&\qquad\qquad\qquad\qquad\qquad\qquad\qquad\
  + \Biggl( \frac{\sigma}{\mu} + \frac{\mu}{\sigma} - 4 \Biggr) \, 
  {\rm Im} \, \Bigl( \frac{\gamma_0}{\omega_0} + i \Bigr)^{-2+d/2} \Biggr] 
  \nonumber \\
  &&\qquad\qquad\! 
  + \lambda \, \frac{\Gamma(3-d/2)}{3 \cdot 2^{d+5} \, \pi^{d/2}} \, 
  \Bigl( \frac{\omega_0}{D_0} \Bigr)^{-1+d/2} \,
  \Biggl[ \Biggl( \frac{\sigma}{\mu} + \frac{\mu}{\sigma} - 4 \Biggr) \, 
  {\rm Re} \, \Bigl( \frac{\gamma_0}{\omega_0} + i \Bigr)^{-3+d/2} \nonumber \\
  &&\qquad\qquad\qquad\qquad\qquad 
  - 3 \, \Biggl( \sqrt{\frac{\sigma}{\mu}} - \sqrt{\frac{\mu}{\sigma}} \, 
  \Biggr) \, {\rm Im} \, \Bigl( \frac{\gamma_0}{\omega_0} + i \Bigr)^{-3+d/2} 
  \Biggr] + O(\lambda^2) \nonumber \\
  &&\quad\ = D_0 + \lambda \, \Bigl( \frac{\omega_0}{D_0} \Bigr)^{-1+d/2} \,
  \Delta {\tilde D}_R + O(\lambda^2) \, .
\label{intdif} 
\end{eqnarray}

Notice that the effective expansion parameter in this perturbation series is 
given by $(\lambda / \omega_0) \, (\omega_0 / D_0)^{d/2}$; accordingly we have 
introduced dimensionless first-order fluctuation corrections 
$\Delta {\tilde \gamma}_R$, $\Delta {\tilde \omega}_R$, and 
$\Delta {\tilde D}_R$.
Naturally, when diffusion is fast compared to the characteristic oscillation 
time scale, the system becomes well-mixed and spatial correlations irrelevant.
Deviations from mean-field theory induced by the fluctuation loops are then
minute.
In dimensions $d < 2$, when we let $\gamma_0 \to 0$, the leading fluctuation 
correction to the oscillation frequency diverges 
$\sim (\omega_0 / \gamma_0)^{1 - d/2}$; it is negative, and symmetric under 
formal rate exchange $\mu \longleftrightarrow \sigma$, c.f. the last term in the 
second line in eq.~(\ref{intfre}).
If we interpret $\gamma_0$ in the above equations as a small, self-consistently
determined damping, these features are in remarkable qualitative agreement with 
our earlier Monte Carlo observations:
Fluctuations and correlations induced by the stochastic reaction processes 
induce a strong downward numerical renormalization of the oscillation 
frequency, with very similar functional dependence on the rates $\mu$ and 
$\sigma$.
Note that $d_c = 2$ can be viewed as (upper) critical dimension for the 
appearance of singular infrared fluctuation contributions (in the limit of 
infinite prey carrying capacity $\rho \to \infty$ or $\gamma_0 \to 0$), which 
resemble dynamic coexistence anomalies induced by Goldstone modes in systems 
with broken continuous order parameter symmetry (see, e.g., 
Ref.~\cite{Tauber92} and references therein).

\begin{figure}[ht]
\begin{center}
\includegraphics[width=10cm]{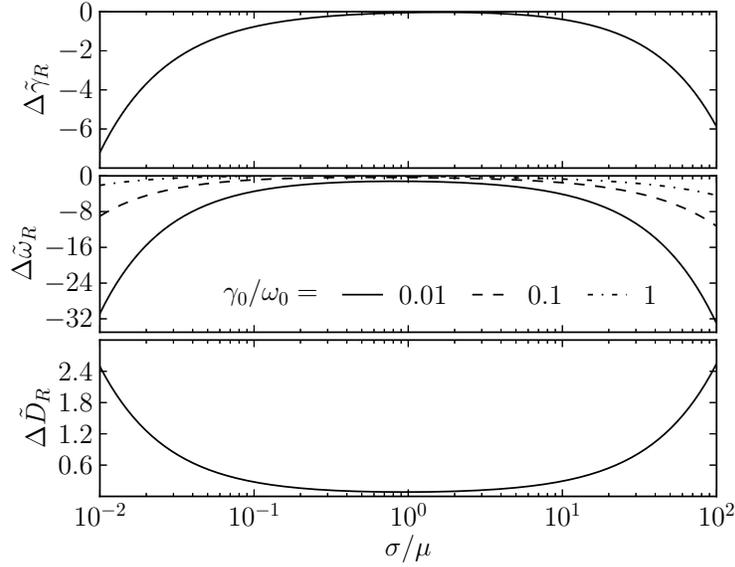}
\end{center}
\caption{Fluctuation contributions to the damping $\Delta {\tilde \gamma}_R$,
  oscillation frequency $\Delta {\tilde \omega}_R$, and diffusivity 
  $\Delta {\tilde D}_R$ in $d = 1$ dimension.
  The fluctuation corrections to the frequency depend crucially on the ratio
  $\gamma_0 / \omega_0$, especially when $\sigma \ll \mu$ or $\sigma \gg \mu$.}
\label{lvmrd1}
\end{figure}
In the following, the expressions (\ref{intdam})--(\ref{intdif}) are evaluated 
at integer dimensions $d = 1,2,3$, and $4$.
In low dimensions, i.e., for $d = 1$ and $d = 2$, the renormalized oscillation 
frequency (\ref{intfre}) becomes singular in the limit $\gamma_0 \to 0$, caused
by the interference of counter-propagating clock- and anti-clockwise internal
modes.
For the renormalized diffusivity $D_R$ and the fluctuation-generated damping 
$\gamma_R$, these infrared singularities cancel out.
In $d = 1$ dimension, the leading terms in $\gamma_0$ are:
\begin{eqnarray}
  &&\gamma_R = \gamma_0 + \frac{\lambda}{8 \, \sqrt{2}} \,    
  \sqrt{\frac{\omega_0}{D_0}} \, \left[ 1 + \frac{3}{4} 
  \left( \sqrt{\frac{\sigma}{\mu}} - \sqrt{\frac{\mu}{\sigma}} \right) 
  - \frac{3}{4} \left( \frac{\sigma}{\mu} + \frac{\mu}{\sigma} \right) \right]
  + O(\lambda^2) \ , \nonumber \\
  && 
\label{rndam1} \\
  &&\omega_R = \omega_0 - \frac{\lambda}{16} \, 
  \frac{\omega_0}{\sqrt{D_0 \, \gamma_0}} \, \left[ 1 + \frac{1}{2} \left(
  \frac{\sigma}{\mu} + \frac{\mu}{\sigma} \right) \right] 
\label{renfr1} \\
  &&\qquad\;\ 
  + \frac{11 \, \lambda}{64\, \sqrt{2}} \, \sqrt{\frac{\omega_0}{D_0}} \, 
  \left[ 1 - \frac{57}{44} \, \sqrt{\frac{\sigma}{\mu}} + \frac{25}{44} \,
  \sqrt{\frac{\mu}{\sigma}} + \frac{1}{44} \left( \frac{\sigma}{\mu} + 
  \frac{\mu}{\sigma} \right) \right] + O(\lambda^2) \ , \nonumber \\
  &&D_R = D_0 + \frac{3\, \lambda}{64\, \sqrt{2}}\, \sqrt{\frac{D_0}{\omega_0}}
  \left[ 1 + \frac{1}{12} \left( \! \sqrt{\frac{\sigma}{\mu}} - 
  \sqrt{\frac{\mu}{\sigma}} \right) + \frac{3}{4} \left( \frac{\sigma}{\mu} + 
  \frac{\mu}{\sigma} \right) \right] \! + O(\lambda^2) \, . \nonumber \\
\label{rndif1} 
\end{eqnarray}
The dimensionless fluctuation corrections $\Delta {\tilde \gamma}_R$,
$\Delta {\tilde \omega}_R$, and $\Delta {\tilde D}_R$, c.f. 
eqs.~(\ref{intdam})--(\ref{intdif}), are plotted in Fig.~\ref{lvmrd1}.
The fluctuation-induced contribution to the damping is always negative, 
indicating the instability towards spatially inhomogeneous structures that 
emerge when $\gamma_0 = \lambda \, |\Delta {\tilde \gamma}_R| \, 
(\omega_0 / D_0)^{d/2} + O(\lambda^2)$. 
The oscillation frequency is renormalized to lower values by the loop 
corrections, with the leading term $\sim \sqrt{\omega_0 / \gamma_0}$ further
amplified when either $\sigma \ll \mu$ or $\sigma \gg \mu$.
Likewise, fluctuations invariably enhance diffusive spreading.
The fluctuation corrections all appear remarkably symmetric with respect to
exchanging $\sigma \longleftrightarrow \mu$, as is evident in Fig.~\ref{lvmrd1} 
with its logarithmic scale for the rate ratio $\sigma / \mu$ by the approximate
mirror symmetry about the $\sigma / \mu = 1$ axis.

\begin{figure}[ht]
\begin{center}
\includegraphics[width=10cm]{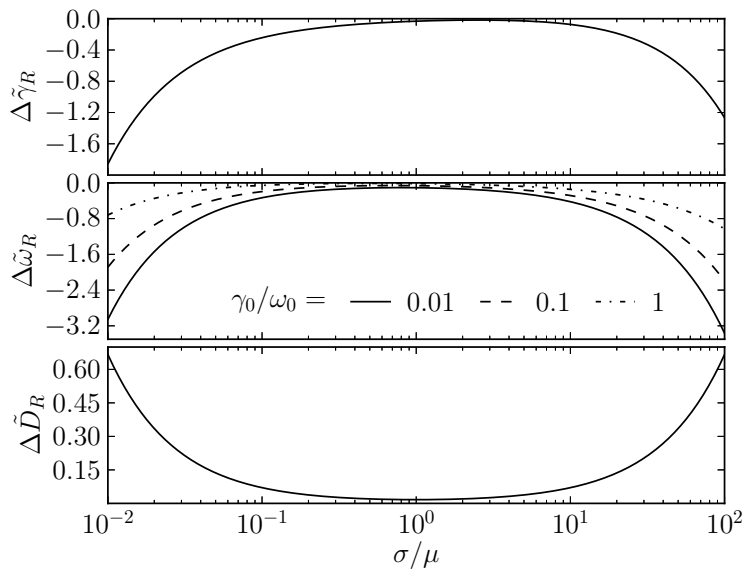}
\end{center}
\caption{Fluctuation contributions to the damping $\Delta {\tilde \gamma}_R$,
  oscillation frequency $\Delta {\tilde \omega}_R$, and diffusivity 
  $\Delta {\tilde D}_R$ in $d = 2$ dimensions.
  As in one dimension, the fluctuation corrections to the frequency strongly 
  depend on the ratio $\gamma_0 / \omega_0$.}
\label{lvmrd2}
\end{figure}
In $d = 2$ dimensions, one gets
\begin{eqnarray}
  &&\gamma_R = \gamma_0 + \frac{\lambda}{64} \, \frac{\omega_0}{D_0} 
  \left[ \frac{6}{\pi} \left( \sqrt{\frac{\sigma}{\mu}} - \sqrt{\frac{\mu}{\sigma}} 
  \right) - \left( \frac{\sigma}{\mu} + \frac{\mu}{\sigma} \right) \right] 
  + O(\lambda^2) \ , 
\label{rndam2} \\
  &&\omega_R = \omega_0 - \frac{\lambda}{32 \, \pi} \, \frac{\omega_0}{D_0} \, 
  \ln\frac{\omega_0}{\gamma_0} \, \cdot \left[ 1 + \frac{1}{2} \left( 
  \frac{\sigma}{\mu} + \frac{\mu}{\sigma} \right) \right] \nonumber \\
  &&\qquad\quad\,\ + \frac{3 \, \lambda}{32 \, \pi} \, \frac{\omega_0}{D_0} \, 
  \left[ 1 - \frac{\pi}{3} \, \sqrt{\frac{\sigma}{\mu}} - \frac{1}{4} \left( 
  \frac{\sigma}{\mu} + \frac{\mu}{\sigma} \right) \right] + O(\lambda^2) \ ,
\label{renfr2} \\
  &&D_R = D_0 + \frac{\lambda}{96\, \pi} \left[ 1 + 2 \left( \frac{\sigma}{\mu}
  + \frac{\mu}{\sigma} \right) \right] + O(\lambda^2) \ ,
\label{rndif2}
\end{eqnarray}
and the fluctuation contributions are depicted in Fig.~\ref{lvmrd2}.
The graphs look remarkably alike to Fig.~\ref{lvmrd1} for $d = 1$, but the 
overall scale of the corrections $\Delta {\tilde \gamma}_R$ and 
$\Delta {\tilde D}_R$ is reduced by a factor $\sim 4$, and for 
$\Delta {\tilde \omega}_R$ by $\sim 10$, with its leading term acquiring just a
logarithmic singularity as $\gamma_0 \to 0$.
Again, the system is rendered unstable against spatio-temporal structures.
According to eq.~(\ref{frtspd}), the fluctuation-enhanced diffusivity suggests 
faster front spreading than predicted by the bare mean-field rates, as indeed
observed in two-dimensional Monte Carlo simulations \cite{Dobramysl08}.
The population oscillation frequency is strongly renormalized downward, with
an approximately equal functional dependence on the rates $\sigma$ and $\mu$;
moreover, the deviations from the mean-field values grow in size as the ratio 
$\sigma / \mu$ is tuned away from unity.
These analytic perturbative one-loop results are in remarkable qualitative 
agreement with the Monte Carlo simulation data for two-dimensional stochastic 
Lotka--Volterra systems, as shown in Fig.~9 in Ref.~\cite{Mobilia07} and 
Fig.~6(b) in Ref.~\cite{Washenberger07}.

\begin{figure}[ht]
\begin{center}
\includegraphics[width=10cm]{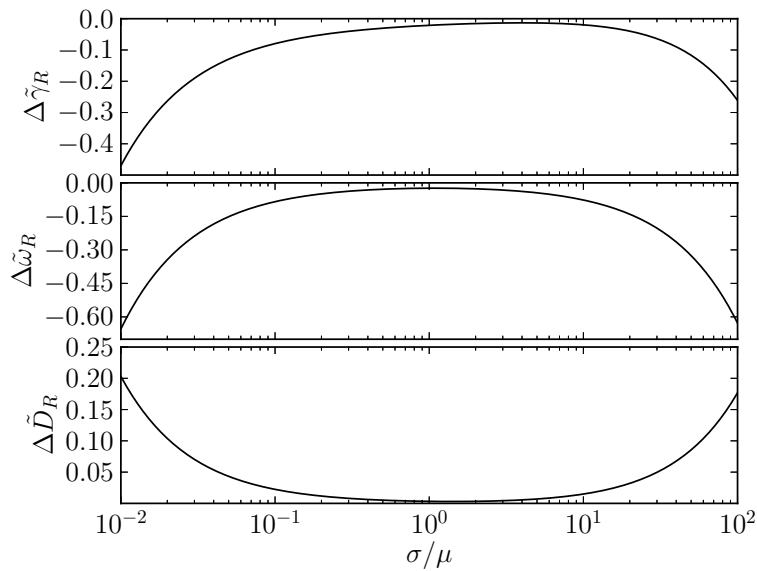}
\end{center}
\caption{Fluctuation contributions to the damping $\Delta {\tilde \gamma}_R$,
  oscillation frequency $\Delta {\tilde \omega}_R$, and diffusivity 
  $\Delta {\tilde D}_R$ in $d = 3$ dimensions.}
\label{lvmrd3}
\end{figure}
In $d = 3$ dimensions, one may safely set the bare damping constant to zero, 
$\gamma_0 \to 0$ (or $\rho \to \infty$) to obtain
\begin{eqnarray}
  &&\gamma_R = \gamma_0 + 
  \frac{\lambda (\omega_0 / D_0)^{3/2}}{16 \, \sqrt{2}\, \pi} \left[ - 1 + 
  \frac{3}{4} \left( \! \sqrt{\frac{\sigma}{\mu}} - \sqrt{\frac{\mu}{\sigma}} 
  \right) - \frac{1}{4} \left( \frac{\sigma}{\mu} + \frac{\mu}{\sigma} \right) 
  \right] \! + O(\lambda^2) \, , \nonumber \\
\label{rndam3} \\
  &&\omega_R = \omega_0 
  + \frac{\lambda (\omega_0 / D_0)^{3/2}}{128 \, \sqrt{2} \, \pi} \left[ 1 - 
  \frac{13}{4} \sqrt{\frac{\sigma}{\mu}} - \frac{19}{4} 
  \sqrt{\frac{\mu}{\sigma}} - \frac{13}{4} \left( \frac{\sigma}{\mu} + 
  \frac{\mu}{\sigma} \right) \right] \! + O(\lambda^2) \, , \nonumber \\
\label{renfr3} \\
  &&D_R = D_0 - \frac{\lambda \sqrt{\omega_0 / D_0}}{384 \, \sqrt{2} \, \pi}
  \left[ 1 + \frac{9}{4} \left( \! \sqrt{\frac{\sigma}{\mu}} - 
  \sqrt{\frac{\mu}{\sigma}} \right) - \frac{13}{4} \left( \frac{\sigma}{\mu} + 
  \frac{\mu}{\sigma} \right) \right] + O(\lambda^2) \, . \nonumber \\
\label{rndif3}
\end{eqnarray}
Figure~\ref{lvmrd3} shows the associated fluctuation corrections 
$\Delta {\tilde \gamma}_R$, $\Delta {\tilde \omega}_R$, and 
$\Delta {\tilde D}_R$, which compared to one and two dimensions are 
considerably reduced in magnitude, but otherwise display quite similar features.

\begin{figure}[t]
\begin{center}
\includegraphics[width=10cm]{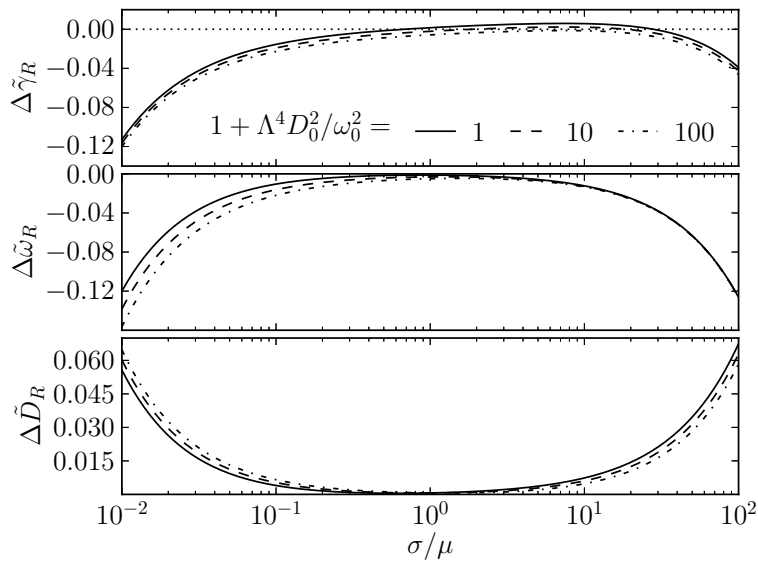}
\end{center}
\caption{Fluctuation contributions to the damping $\Delta {\tilde \gamma}_R$,
  oscillation frequency $\Delta {\tilde \omega}_R$, and diffusivity 
  $\Delta {\tilde D}_R$ in $d = 4$ dimensions.
  Notice the weak logarithmic dependence on the ultraviolet cut-off $\Lambda$.}
\label{lvmrd4}
\end{figure}

In higher dimensions $d \geq 4$, the fluctuation corrections become formally
ultraviolet-divergent, and thus a finite cut-off $\Lambda$ in momentum space 
must be implemented; e.g., in $d = 4$ dimensions:
\begin{eqnarray}
  &&\gamma_R = \gamma_0 + \frac{\lambda}{32\, \pi^2} \left( 
  \frac{\omega_0}{D_0} \right)^2 \Biggl[ 1 - \frac{1}{2} \, \ln \left( 1 + 
  \frac{\Lambda^4}{\omega_0^2/D_0^2} \right) + \frac{3 \, \pi}{8} \left( 
  \sqrt{\frac{\sigma}{\mu}} - \sqrt{\frac{\mu}{\sigma}} \right) \nonumber \\
  &&\qquad\qquad\qquad\qquad\quad - \frac{1}{4} \left( \frac{\sigma}{\mu} 
  + \frac{\mu}{\sigma} \right) \Biggr] + O(\lambda^2) \ , 
\label{rndam4} \\
  &&\omega_R = \omega_0 + \frac{\lambda}{256 \, \pi} \left( 
  \frac{\omega_0}{D_0} \right)^2 \Biggl[ 1 - \frac{2}{\pi} \, 
  \sqrt{\frac{\mu}{\sigma}} \, \ln \left( 1 + 
  \frac{\Lambda^4}{\omega_0^2/D_0^2} \right) \nonumber \\
  &&\qquad\qquad\qquad\qquad\quad - \frac{5}{2 \, \pi^2} \left( 
  \sqrt{\frac{\sigma}{\mu}} - \sqrt{\frac{\mu}{\sigma}} \right) - \left( 
  \frac{\sigma}{\mu} + \frac{\mu}{\sigma} \right) \Biggr] + O(\lambda^2) \, ,
\label{renfr4} \\
  &&D_R = D_0 - \frac{\lambda}{512 \, \pi} \, \frac{\omega_0}{D_0} \, \Biggl[ 1
  + \frac{1}{\pi} \left( \sqrt{\frac{\sigma}{\mu}} - \sqrt{\frac{\mu}{\sigma}} 
  \right) \, \ln \left( 1 + \frac{\Lambda^4}{\omega_0^2/D_0^2} \right) 
  \nonumber \\
  &&\qquad\qquad\qquad\qquad\quad - \frac{3}{\pi} \left( 
  \sqrt{\frac{\sigma}{\mu}} - \sqrt{\frac{\mu}{\sigma}} \right) - \left( 
  \frac{\sigma}{\mu} + \frac{\mu}{\sigma} \right) \Biggr] + O(\lambda^2) \ . 
\label{rndif4}
\end{eqnarray}
As Fig.~\ref{lvmrd4} demonstrates, the cut-off dependence in the loop 
corrections is rather weak in $d = 4$ dimensions.
For low cut-off values, the fluctuation contributions to the damping appear 
positive in the approximate interval $1 \leq \sigma / \mu \leq 30$, but turn
negative in the continuum limit of large $\Lambda$, still signaling instability
with respect to structure formation.
The typical values of $\Delta {\tilde \gamma}_R$, $\Delta {\tilde D}_R$, and
$\Delta {\tilde \omega}_R$ are all diminished by factors $\sim 4 \ldots 5$ as 
compared to $d = 3$; the cut-off dependence in the renormalized frequency only 
becomes noticeable for $\sigma / \mu \ll 1$, see eq.~(\ref{renfr4}).

\section{Conclusion and outlook}

This paper describes in some detail how the stochastic kinetics of spatially 
extended predator-prey systems of the Lotka--Volterra type, as encoded through
a classical master equation, can be mapped onto a continuum field theory 
representation, while faithfully preserving the internal demographic and 
reaction noise and the ensuing correlations.
The connection of the more microscopic Doi--Peliti field theory action with a
mesoscopic description in terms of coupled Langevin equations was pointed out,
and the associated white noise correlations were systematically derived.
The continuum representation was then employed to demonstrate that the predator
extinction transition, induced by a finite prey carrying capacity, is indeed
governed by the universal scaling exponents of critical directed percolation,
as one would generically expect for such a nonequilibrium phase transition from
an active to an absorbing state.

After a brief review of the most striking features of stochastic predator-prey 
models in the species coexistence phase, the Doi--Peliti field theory 
representation and a first-order perturbation expansion with respect to the
nonlinear predation rate, in the limit of large prey carrying capacity, were 
employed to qualitatively and semi-quantitatively confirm crucial salient 
observations from Monte Carlo simulations on regular lattices:
(i) Spatial predator-prey systems in the species coexistence phase are 
generically characterized by the emergence of fluctuating spatial structures,
namely continually expanding and merging activity fronts.
(ii) The recurring passages of population waves locally incite persistent 
density oscillations for both predators and prey.
(iii) Fluctuations in the two-species coexistence phase are remarkably and 
quite unusually strong; as compared with the (linearized) mean-field 
prediction, they considerably renormalize the oscillation frequency, especially
in $d \leq 2$ dimensions.
Explicit analytical results for the fluctuation-induced damping, and the 
renormalized oscillation frequency and diffusion coefficient were provided 
to one-loop order.
They showed that (iv) the leading fluctuation contribution to the frequency is 
negative, and symmetric in its functional dependence on the rates $\sigma$ and
$\mu$; and (v) the diffusivity is invariably renormalized upward, implying
faster front propagation speeds as compared to the mean-field approximation.

An important open question is which of the numerous standard mathematical 
models in ecology, population dynamics, and chemical kinetics, many of which 
are frequently analyzed merely on the level of mean-field rate equations, are 
similarly strongly affected by stochastic fluctuations and intrinsic 
correlations.
Remarkably, and perhaps counter-intuitively, Monte Carlo simulations of 
stochastic spatial variants of cyclic three-species predator-prey systems,
namely both spatial rock-paper-scissors games (with conserved total population)
and the May--Leonard model (which displays no global conservation law) do not
reveal noticeable fluctuation effects, see Refs.~\cite{He10, He11} (and further
references therein).
Apparently, the mechanism causing strong fluctuations in the spatial stochastic
two-species Lotka--Volterra system, namely the destructive interference of 
counter-propagating internal modes, is conspicuously absent in extensions to 
additional participating species.
This fact becomes even more puzzling as the stochastic cyclic 
rock-paper-scissors model has been shown to reduce to the stochastic and 
strongly fluctuating Lotka--Volterra system in a highly asymmetric rate limit 
where a single species becomes abundant \cite{He12}.
A careful field-theoretic analysis based on the Doi--Peliti representation of
the corresponding stochastic master equation should be capable to shed light on
this issue, and hopefully explain this important distinction between apparently
closely related population dynamics or reaction-diffusion models.

\ack
The author is indebted to Ulrich Dobramysl for his assistance with 
Figs.~\ref{lvmrd1}--\ref{lvmrd4}, and gladly acknowledges helpful discussions 
with him, Qian He, Swapnil Jawkar, Mauro Mobilia, Michel Pleimling, 
Beate Schmittmann, Siddharth Venkat, and Royce Zia.

\appendix

\section{Field theory and counter-terms for finite carrying capacity}
\label{fulfth}

In this appendix, we write down the explicit field theory for finite prey
carrying capacity $\rho$, and sketch the evaluation of the associated 
counter-terms $A_c$ and $B_c$ to first order in the predation rate $\lambda$. 
Upon expressing (\ref{sffsr}) in terms of the fields ${\widetilde \varphi}_\pm$
and $\varphi_\pm$ by means of eqs.~(\ref{cphph}), one obtains the source terms
\begin{eqnarray}
  &&\!\!\!\!\!\!\!\!\!\!\!\!\!\!\!\!\!\!\!\!\!\!\!\!\!\!\!\!\!\!\!\!\!\! 
  S_s[{\widetilde \varphi}_\pm;\varphi_\pm] = \int \! d^dx \! \int \!\! dt \, 
  \Biggl[ \, \sqrt{\frac{\mu}{2}} \, \frac{\sigma}{i \omega_0 \, \lambda} \, 
  \Biggl( \biggl[ (i \omega_0 + \gamma_0) \, (1 + B_c) \, \biggl( A_c + 
  \frac{\mu}{\rho \, \lambda} \, B_c \biggr) - \mu B_c \times \nonumber \\
  &&\!\!\!\!\!\!\!\!\!\!\!\!\!\!\!\!\!\!\!\!\!\!\!\!\!\!\!\!\!\!\!\!\!
  \biggl( 1 - \frac{\mu}{\rho \, \lambda} + A_c \biggr) \biggr] \, 
  {\widetilde \varphi}_+ + \biggl[ (i \omega_0 - \gamma_0) \, (1 + B_c) \, 
  \biggl( A_c + \frac{\mu}{\rho \, \lambda} \, B_c \biggr) + \mu B_c \, \biggl(
  1 - \frac{\mu}{\rho \, \lambda} + A_c \biggr) \biggr] \, 
  {\widetilde \varphi}_- \! \Biggr) \nonumber \\
  &&\!\!\!\!\!\!\!\!\!\!\!\!\!\!\!\!\!\!\!\!\!\!\!\!\!\!\!\!\! 
  + \frac{\sigma (1 + B_c)}{2 \omega_0^2 \, \lambda} \Biggl( \biggl[ 
  (i \omega_0 + \gamma_0)^2 \biggl( 1 - \frac{\alpha \, \mu}{\rho \, \lambda} 
  \, (1 + B_c) \biggr) - \mu (i \omega_0 + \gamma_0 - \mu) \biggl( 1 - 
  \frac{\mu}{\rho \, \lambda} + A_c \biggr) \biggr] \, {\widetilde \varphi}_+^2
  \nonumber \\
  &&\quad - 2 \, \biggl[ (\omega_0^2 + \gamma_0^2) \, \biggl( 1 - 
  \frac{\alpha \, \mu}{\rho \, \lambda} \, (1 + B_c) \biggr) - \mu \, 
  (\gamma_0 - \mu) \biggl( 1 - \frac{\mu}{\rho \, \lambda} + A_c \biggr) 
  \biggr] \, {\widetilde \varphi}_+ \, {\widetilde \varphi}_- \nonumber \\
  &&\!\! + \biggl[ (i \omega_0 - \gamma_0)^2 \biggl( 1 - 
  \frac{\alpha \mu}{\rho \, \lambda} \, (1 + B_c) \biggr) + \mu (i \omega_0
  - \gamma_0 + \mu) \biggl( 1 - \frac{\mu}{\rho \, \lambda} + A_c \biggr) 
  \biggr] \, {\widetilde \varphi}_-^2 \Biggr) \nonumber \\
  &&\!\! - (\alpha - 1) \, \sqrt{\frac{\mu}{2}} \, 
  \frac{\sigma \, (1 + B_c)^2}{2 i \, \omega_0^3 \, \rho \, \lambda^2} \, 
  \Bigl[ (i \omega_0 + \gamma_0) \, {\widetilde \varphi}_+ 
  + (i \omega_0 - \gamma_0) \, {\widetilde \varphi}_- \Bigr]^3 \Biggr] \, . 
\label{sphisr} 
\end{eqnarray}
Note that the cubic source contributions are absent if $\alpha = 1$. 
The nonlinear action (\ref{sffnl}) yields the three-point vertices
\begin{eqnarray}
  &&\!\!\!\!\!\!\!\!\!\!\!\!\!\! S_v[{\widetilde \varphi}_\pm;\varphi_\pm] = - 
  \frac{1}{2 \sqrt{2 \mu} \, i \omega_0^3} \, \int \! d^dx \! \int \!\! dt \, 
  \Biggl[ \Biggl( (i \omega_0 + \gamma_0 - \mu) \Bigl[ (i \omega_0 - \gamma_0)
  \, \mu \, (1 + B_c) \nonumber \\
  &&\qquad\quad + (\omega_0^2 + \gamma_0^2 + \mu \, \sigma \, A_c) \Bigr] 
  - (i \omega_0 + \gamma_0)^2 \, \sigma \Bigl[ 1 - 
  \frac{2 \alpha \, \mu}{\rho \, \lambda} \, (1 + B_c) \Bigr] \Biggr) \, 
  {\widetilde \varphi}_+^2 \, \varphi_+ \nonumber \\
  &&\qquad\quad + \Biggl( (i \omega_0 + \gamma_0 - \mu) \Bigl[ (i \omega_0 
  + \gamma_0) \, \mu \, (1 + B_c) - (\omega_0^2 + \gamma_0^2 + \mu \, \sigma 
  \, A_c) \Bigr] \nonumber \\ 
  &&\qquad\qquad\qquad\qquad\qquad\qquad + (i \omega_0 + \gamma_0)^2 \, 
  \sigma \Bigl[ 1 - \frac{2 \alpha \, \mu}{\rho \, \lambda} \, (1 + B_c) \Bigr] 
  \Biggr) \, {\widetilde \varphi}_+^2 \, \varphi_- \nonumber \\
  &&\qquad\quad - 2 \Biggl( (\gamma_0 - \mu) \Bigl[ (i \omega_0 - \gamma_0) \,
  \mu \, (1 + B_c) + (\omega_0^2 + \gamma_0^2 + \mu \, \sigma \, A_c) \Bigr] 
  \nonumber \\ 
  &&\qquad\qquad\qquad\qquad\qquad\quad - (\omega_0^2 + \gamma_0^2) \, 
  \sigma \Bigl[ 1 - \frac{2 \alpha \, \mu}{\rho \, \lambda} \, (1 + B_c) \Bigr] 
  \Biggr) \, {\widetilde \varphi}_+ \, {\widetilde \varphi}_- \, \varphi_+ 
  \nonumber \\  
  &&\qquad\quad - 2 \Biggl( (\gamma_0 - \mu) \Bigl[ (i \omega_0 + \gamma_0) \, 
  \mu \, (1 + B_c) - (\omega_0^2 + \gamma_0^2 + \mu \, \sigma \, A_c) \Bigr] 
  \nonumber \\ 
  &&\qquad\qquad\qquad\qquad\qquad\quad + (\omega_0^2 + \gamma_0^2) \, 
  \sigma \Bigl[ 1 - \frac{2 \alpha \, \mu}{\rho \, \lambda} \, (1 + B_c) \Bigr] 
  \Biggr) \, {\widetilde \varphi}_+ \, {\widetilde \varphi}_- \, \varphi_- 
  \nonumber \\
  &&\qquad\quad - \Biggl( (i \omega_0 - \gamma_0 + \mu) \Bigl[ (i \omega_0 
  - \gamma_0) \, \mu \, (1 + B_c) + (\omega_0^2 + \gamma_0^2 + \mu \, \sigma 
  \, A_c) \Bigr] \nonumber \\ 
  &&\qquad\qquad\qquad\qquad\qquad\qquad + (i \omega_0 - \gamma_0)^2 \, 
  \sigma \Bigl[ 1 - \frac{2 \alpha \, \mu}{\rho \, \lambda} \, (1 + B_c) \Bigr] 
  \Biggr) \, {\widetilde \varphi}_-^2 \, \varphi_+ \nonumber \\
  &&\qquad\quad - \Biggl( (i \omega_0 - \gamma_0 + \mu) \Bigl[ (i \omega_0 
  + \gamma_0) \, \mu \, (1 + B_c) - (\omega_0^2 + \gamma_0^2 + \mu \, \sigma
  \, A_c) \Bigr] \nonumber \\ 
  &&\qquad\qquad\qquad\qquad\qquad\qquad - (i \omega_0 - \gamma_0)^2 \, 
  \sigma \Bigl[ 1 - \frac{2 \alpha \, \mu}{\rho \, \lambda} \, (1 + B_c) \Bigr] 
  \Biggr) \, {\widetilde \varphi}_-^2 \, \varphi_- \nonumber \\
  &&\qquad\qquad\qquad + \Bigl[ \lambda \, (i \omega_0 + \gamma_0 - \mu) 
  (i \omega_0 - \gamma_0) + \frac{\mu \, \sigma}{\rho} \, (i \omega_0 
  + \gamma_0) \Bigr] \, {\widetilde \varphi}_+ \, \varphi_+^2 \nonumber \\
  &&\qquad\qquad\qquad + 2 \Bigl[ \lambda \, \gamma_0 \,  (i \omega_0 
  + \gamma_0 - \mu) - \frac{\mu \, \sigma}{\rho} \, (i \omega_0 + \gamma_0) 
  \Bigr] \, {\widetilde \varphi}_+ \, \varphi_+ \, \varphi_- \nonumber \\
  &&\qquad\qquad\qquad - (i \omega_0 + \gamma_0) \Bigl[ \lambda \, (i \omega_0 
  + \gamma_0 - \mu) - \frac{\mu \, \sigma}{\rho} \Bigr] \, 
  {\widetilde \varphi}_+ \, \varphi_-^2 
\label{nlver3} \\
  &&\qquad\qquad\qquad + (i \omega_0 - \gamma_0) \Bigl[ \lambda \, (i \omega_0 
  - \gamma_0 + \mu) + \frac{\mu \, \sigma}{\rho} \Bigr] \,
  {\widetilde \varphi}_- \, \varphi_+^2 \nonumber \\
  &&\qquad\qquad\qquad + 2 \Bigl[ \lambda \, \gamma_0 \, (i \omega_0 
  - \gamma_0 + \mu) - \frac{\mu \, \sigma}{\rho} \, (i \omega_0 - \gamma_0) 
  \Bigr] \, {\widetilde \varphi}_- \, \varphi_+ \, \varphi_- \nonumber \\
  &&\qquad\qquad\qquad - \Bigl[ \lambda \, (i \omega_0 - \gamma_0 + \mu) 
  (i \omega_0 + \gamma_0) - \frac{\mu \, \sigma}{\rho} \, (i \omega_0 
  - \gamma_0) \Bigr] \, {\widetilde \varphi}_- \, \varphi_-^2 \Biggr] \ . 
  \nonumber
\end{eqnarray}
In addition, there are four- and five-point vertices (the latter arise only for
$\alpha = 2$):
\begin{eqnarray}
  &&\!\!\!\!\!\!\!\!\!\!\!\!\!\!\!\!\!\!\!\!\!\!\!\!\!\!\!\!\!\!\!
  S_v'[{\widetilde \varphi}_\pm;\varphi_\pm] = \int \! d^dx \! \int \!\! dt 
  \Biggl[ (\alpha - 1) \frac{\sigma (1 + B_c)}{2 \,\omega_0^4 \, \rho \, 
  \lambda} \, \Bigl[ (i \omega_0 + \gamma_0) \, {\widetilde \varphi}_+ 
  + (i \omega_0 - \gamma_0) \, {\widetilde \varphi}_- \Bigr]^3  
  (\varphi_+ - \varphi_-) \nonumber \\
  &&+ \frac{\lambda}{4 \, \omega_0^4} \, \Bigl[ (i \omega_0 + \gamma_0 - \mu) 
  \, {\widetilde \varphi}_+^2 - 2 (\gamma_0 - \mu) \, {\widetilde \varphi}_+ \,
  {\widetilde \varphi}_- - (i \omega_0 - \gamma_0 + \mu) \, 
  {\widetilde \varphi}_-^2 \Bigr] \times \nonumber \\
  &&\qquad\qquad\qquad \Bigl[ (i \omega_0 - \gamma_0) \, \varphi_+^2 + 
  2 \gamma_0 \, \varphi_+ \, \varphi_- - (i \omega_0 + \gamma_0) \, \varphi_-^2
  \Bigr] \nonumber \\
  &&+ \frac{\alpha \, \sigma}{4 \, \omega_0^4 \, \rho} \, \Bigl[ 
  (i \omega_0 + \gamma_0) \, {\widetilde \varphi}_+ + (i \omega_0 - \gamma_0) 
  \, {\widetilde \varphi}_- \Bigr]^2 \, (\varphi_+ - \varphi_-)^2
\label{vert45} \\
  &&+ (\alpha - 1) \, \frac{\sigma}{4 \sqrt{2 \mu} \, i \omega_0^5 \, \rho} \,
  \Bigl[ (i \omega_0 + \gamma_0) \, {\widetilde \varphi}_+ 
  + (i \omega_0 - \gamma_0) \, {\widetilde \varphi}_- \Bigr]^3 \, 
  (\varphi_+ - \varphi_-)^2 \Biggr] \ . \nonumber 
\end{eqnarray}
However, these do not enter the one-loop analysis, but only contribute to 
higher orders in the perturbation expansion.

\begin{figure}[ht]
\begin{center}
\includegraphics[width=13cm]{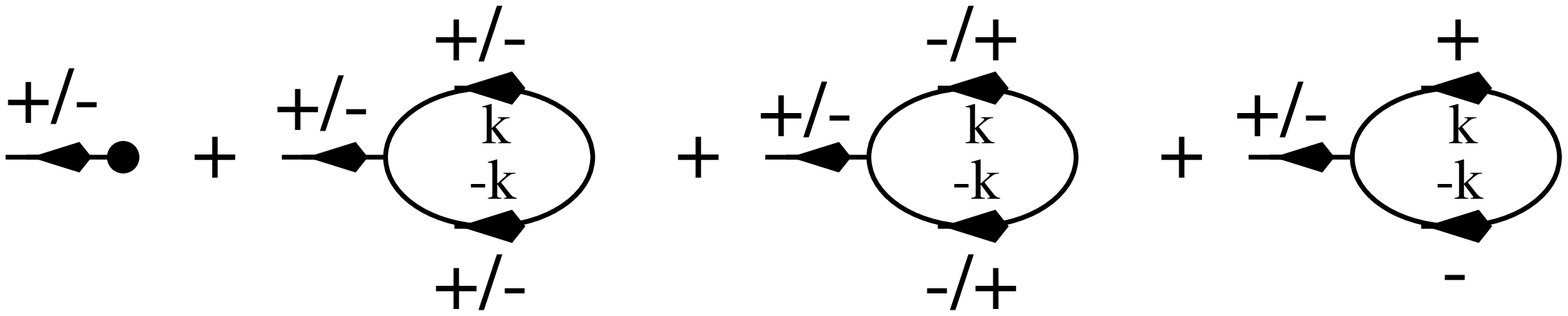}
\end{center}
\caption{Feynman graphs for $\langle \varphi_\pm \rangle$ up to one-loop order
  in the full field theory.}
\label{lvfiga}
\end{figure}
Naturally, with these many contributions in the action, any subsequent
perturbative calculation becomes quite elaborate and lengthy, as will next be
demonstrated by computing the counter-terms $A_c$ and $B_c$ in the full theory.
The contributing Feynman graphs up to one-loop order are depicted in 
Fig.~\ref{lvfiga}.
The associated analytic expressions for the expectation values
$\langle \varphi_\pm \rangle$ become to first order in $\lambda$:
\begin{eqnarray}
  &&\!\!\!\!\!\!\!0 = \langle \varphi_\pm \rangle = (\pm i \omega_0 + \gamma_0)
  \, \Bigl( A_c + \frac{\mu}{\rho \, \lambda} \, B_c \Bigr) - \mu \, \Bigl( 
  1 - \frac{\mu}{\rho \, \lambda} \Bigr) B_c \nonumber \\
  &&\!\!\!\! + \frac{1 - \mu / \rho \, \lambda}{4 \, \omega_0^2 \, \mu} \, 
  \Biggl( \Bigl[ \lambda \, (\pm i \omega_0 + \gamma_0 - \mu) \, 
  (\mp i \omega_0 + \gamma_0) - (\pm i \omega_0 + \gamma_0) \,
  \frac{\mu \, \sigma}{\rho} \Bigr] \nonumber \\ 
  &&\qquad\qquad\quad \times \Bigl[ (\pm i \omega_0 + \gamma_0)^2 - \mu 
  \, (\pm i \omega_0 + \gamma_0 - \mu) \Bigr] \int \! 
  \frac{(2\pi)^{-d} \, d^dk}{\pm i \omega_0 + \gamma_0 + D_0 k^2} \nonumber \\
  &&\qquad\qquad + (\pm i \omega_0 + \gamma_0) \, \Bigl[ \lambda \, 
  (\pm i \omega_0 + \gamma_0 - \mu) - \frac{\mu \, \sigma}{\rho} \Bigr] 
  \nonumber \\ 
  &&\qquad\qquad\quad \times \Bigl[ (\mp i \omega_0 + \gamma_0)^2
  - \mu \, (\mp i \omega_0 + \gamma_0 - \mu) \Bigr] \int \! 
  \frac{(2\pi)^{-d} \, d^dk}{\mp i \omega_0 + \gamma_0 + D_0 k^2} \nonumber \\
  &&\qquad\qquad - 2 \, \Bigl[ \lambda \, \gamma_0 \, 
  (\pm i \omega_0 + \gamma_0 - \mu) - (\pm i \omega_0 + \gamma_0) \,
  \frac{\mu \, \sigma}{\rho} \Bigr] \nonumber \\
  &&\qquad\qquad\qquad\qquad\qquad \times \Bigl[ \omega_0^2 + \gamma_0^2 
  - \mu \, (\gamma_0 - \mu) \Bigr] \int \! 
  \frac{(2\pi)^{-d} \, d^dk}{\gamma_0 + D_0 k^2} \Biggr)
\label{fltexp} \\
  &&\!\!\!\! - \frac{\alpha - 1}{4 \, \omega_0^2 \, \rho \, \lambda} \, 
  \Biggl( \Bigl[ \lambda \, (\pm i \omega_0 + \gamma_0 - \mu) \, 
  (\mp i \omega_0 + \gamma_0) - (\pm i \omega_0 + \gamma_0) \, 
  \frac{\mu \, \sigma}{\rho} \Bigr] \nonumber \\
  &&\qquad\qquad\qquad\qquad\qquad\qquad\qquad\ \times 
  (\pm i \omega_0 + \gamma_0)^2 \int \! 
  \frac{(2\pi)^{-d} \, d^dk}{\pm i \omega_0 + \gamma_0 + D_0 k^2} \nonumber \\
  &&+ \Bigl[ \lambda \, (\pm i \omega_0 + \gamma_0 - \mu) -
  \frac{\mu \, \sigma}{\rho} \Bigr] \, (\omega_0^2 + \gamma_0^2) \, 
  (\mp i \omega_0 + \gamma_0) \int \! 
  \frac{(2\pi)^{-d} \, d^dk}{\mp i \omega_0 + \gamma_0 + D_0 k^2} \nonumber \\
  &&\quad - 2 \Bigl[ \lambda \, \gamma_0 \, 
  (\pm i \omega_0 + \gamma_0 - \mu) - (\pm i \omega_0 + \gamma_0) 
  \frac{\mu \, \sigma}{\rho} \Bigr] (\omega_0^2 + \gamma_0^2)
  \int \! \frac{(2\pi)^{-d} \, d^dk}{\gamma_0 + D_0 k^2}
  \Biggr) \, . \nonumber
\end{eqnarray}
Because of the fundamental symmetry (\ref{vfsym}), separating (\ref{fltexp}) 
into its real and imaginary parts yields only two coupled linear equations for 
$A_c$ and $B_c$.
By means of straightforward (but tedious) algebra one finally obtains
\begin{eqnarray}
  &&\!\!\!\!\!\!\!\!\!\!\!\!\!\!\!\!\!\!\!\!\!\! A_c = 
  - \frac{1}{2 \, \omega_0^2 \, \mu} \, \Biggl[ \Biggl( \lambda \, \Bigl[
  (\omega_0^2 + \gamma_0^2) \, (\gamma_0 - \mu) + \gamma_0 \, \mu^2 \Bigr] 
  + \frac{\mu \, \sigma}{\rho} \, \Bigl[ \omega_0^2 - \gamma_0^2 + \mu \, 
  (\gamma_0 - \mu) \Bigr] \nonumber \\
  &&\!\! - \frac{(\alpha - 1) \, \mu}{\rho \, \lambda - \mu} \, \Bigl[ 
  \lambda \, (\omega_0^2 + \gamma_0^2) \, \gamma_0 
  + \frac{\mu \, \sigma}{\rho} \, (\omega_0^2 - \gamma_0^2) \Bigr] \Biggr) \int\! 
  \frac{d^dk}{(2 \pi)^d} \, \frac{\gamma_0 + D_0 k^2}{\omega_0^2 
  + (\gamma_0 + D_0 k^2)^2} \nonumber \\
  &&\qquad\ + \omega_0^2 \, \Biggl( \lambda\, (\omega_0^2 + \gamma_0^2 - 
  \mu^2) - \frac{\mu \, \sigma}{\rho} \, (2 \, \gamma_0 - \mu)
\label{flcnta} \\
  &&\qquad\qquad\quad\ - \frac{(\alpha - 1) \, \mu}{\rho \, \lambda - \mu} \, 
  \Bigl[ \lambda \, (\omega_0^2 + \gamma_0^2) - 2 \, \frac{\mu \, \sigma}{\rho}
  \, \gamma_0 \Bigr] \Biggr) \int \! \frac{(2 \pi)^{-d} \, d^dk}{\omega_0^2 + 
  (\gamma_0 + D_0 k^2)^2} \nonumber \\
  &&\!\!\!\!\!\!\!\!\! - \Bigl( \lambda \, \gamma_0 
  - \frac{\mu \, \sigma}{\rho} \Bigr) \Bigl( \omega_0^2 + \gamma_0^2 - \mu \, 
  (\gamma_0 - \mu) - \frac{(\alpha - 1)\, \mu}{\rho \, \lambda - \mu} \, 
  (\omega_0^2 + \gamma_0^2) \Bigr) \int \! 
  \frac{(2 \pi)^{-d} \, d^dk}{\gamma_0 + D_0 k^2} \, \Biggr] \nonumber \\
  &&\!\!\!\!\!\!\!\!\!\!\!\!\!\!\!\!\!\!\! 
  + \frac{1}{2 \, \omega_0^2 \, \rho \, \lambda} \, \Biggl[ \Biggl( 2 \, 
  \lambda \, \Bigl[ \omega_0^2 \, (\gamma_0 - \mu) + \gamma_0 \, [\gamma_0^2 
  - \mu \, (\gamma_0 - \mu)] + \frac{\mu \, \sigma}{\rho} \, \Bigl[ \omega_0^2
  - \gamma_0^2 + \mu \, (\gamma_0 - \mu) \Bigr] \nonumber \\
  &&\!\!\!\!\!\! - \frac{(\alpha - 1) \, \mu}{\rho \, \lambda - \mu} \, \Bigl[ 
  2 \, \lambda \, (\omega_0^2 + \gamma_0^2) \, \gamma_0 
  + \frac{\mu \, \sigma}{\rho} \, (\omega_0^2 - \gamma_0^2) \Bigr] \Biggr) 
  \int\! \frac{d^dk}{(2 \pi)^d} \, \frac{\gamma_0 + D_0 k^2}{\omega_0^2 +
  (\gamma_0 + D_0 k^2)^2} \nonumber \\
  &&\qquad\ + \omega_0^2 \, \Biggl( 2 \, \lambda \, 
  (\omega_0^2 + \gamma_0^2 - \mu^2) - \frac{\mu \, \sigma}{\rho} \, 
  (2 \, \gamma_0 - \mu) \nonumber \\
  &&\qquad\qquad\quad\ - \frac{2 \, (\alpha - 1) \, \mu}{\rho \, \lambda - \mu}
  \, \Bigl[ \lambda \, (\omega_0^2 + \gamma_0^2) - \frac{\mu \, \sigma}{\rho} 
  \, \gamma_0 \Bigr] \Biggr) \int \! \frac{(2 \pi)^{-d} \, d^dk}{\omega_0^2 + 
  (\gamma_0 + D_0 k^2)^2} \nonumber \\
  &&\!\!\!\!\!\!\!\!\!\!\!\!\!\!\! - \Bigl( 2 \, \lambda \, \gamma_0 
  - \frac{\mu \, \sigma}{\rho} \Bigr) \Bigl( \omega_0^2 + \gamma_0^2 - \mu \, 
  (\gamma_0 - \mu) - \frac{(\alpha - 1)\, \mu}{\rho \, \lambda - \mu} \, 
  (\omega_0^2 + \gamma_0^2) \Bigr) \int \! 
  \frac{(2 \pi)^{-d} \, d^dk}{\gamma_0 + D_0 k^2} \, \Biggr] \, , \nonumber
\end{eqnarray}
and
\begin{eqnarray}
  &&B_c = - \frac{\lambda}{2 \, \omega_0^2 \, \mu} \, \Biggl( \Bigl[ \omega_0^2
  \, (\gamma_0 - \mu) + \gamma_0 \, [\gamma_0^2 - \mu \, (\gamma_0 - \mu)] 
\label{ftcntb} \\
  &&\qquad\qquad\qquad\qquad - \frac{(\alpha - 1)\, \mu}{\rho \, \lambda - \mu}
  \, (\omega_0^2 + \gamma_0^2) \, \gamma_0 \Bigr] \int\! 
  \frac{d^dk}{(2 \pi)^d} \, \frac{\gamma_0 + D_0 k^2}{\omega_0^2 + 
  (\gamma_0 + D_0 k^2)^2} \nonumber \\
  &&\qquad\quad\ + \omega_0^2 \, \Bigl[ \omega_0^2 + \gamma_0^2 - \mu^2 
  - \frac{(\alpha - 1) \, \mu}{\rho \, \lambda - \mu} \, 
  (\omega_0^2 + \gamma_0^2) \Bigr] \int \! \frac{(2 \pi)^{-d} \, d^dk}
  {\omega_0^2 + (\gamma_0 + D_0 k^2)^2} \nonumber \\
  &&\qquad\quad - \gamma_0 \Bigl[ \omega_0^2 + \gamma_0^2 - \mu \, 
  (\gamma_0 - \mu) - \frac{(\alpha - 1) \, \mu}{\rho \, \lambda - \mu} \, 
  (\omega_0^2 + \gamma_0^2) \Bigr] 
  \int \! \frac{(2 \pi)^{-d} \, d^dk}{\gamma_0 + D_0 k^2} \Biggr)\, . \nonumber
\end{eqnarray}
In the large prey carrying capacity limit $\rho \to \infty$ with 
$\gamma_0 \to 0$, these expression coincide and reduce to eq.~(\ref{cnt1lp}).

\section{Evaluation of the one-loop vertex function}
\label{gameva}

Next we provide some intermediate steps andadditional technical details for
the evaluation of the propagator self-energy of vertex function 
$\Gamma_{\pm;\pm}({\vec q},\omega)$ that results in the renormalized damping
coefficient $\gamma_R$, frequency $\omega_R$, and diffusivity $D_R$.

Collecting and rearranging the one-loop contributions in eq.~(\ref{prp1lp}), one 
arrives at
\begin{eqnarray}
  &&\ \ {\rm Re} \, \Gamma_{\pm ; \pm}^{(1)}(0,0) = 
  + \lambda \, \frac{\sigma - 3 \mu}{8} \int \! \frac{d^dk}{(2 \pi)^d} \ 
  \frac{1}{\omega_0^2 + (\gamma_0 + D_0 \, k^2)^2} \nonumber \\
  &&\qquad\qquad\qquad\quad - \lambda \, \frac{(\sigma + \mu)^2}{8} \int \! 
  \frac{d^dk}{(2 \pi)^d} \ \frac{1}{\gamma_0 + D_0 \, k^2} \ 
  \frac{1}{\omega_0^2 + (\gamma_0 + D_0 \, k^2)^2}  \nonumber \\
  &&\qquad\qquad\qquad\quad - \lambda \, 
  \frac{\sigma^2 - 4 \, \sigma \mu + \mu^2}{4} \int \! \frac{d^dk}{(2 \pi)^d} 
  \ \frac{\gamma_0 + D_0 \, k^2}{[\omega_0^2 + (\gamma_0 + D_0 \, k^2)^2]^2}  
  \nonumber \\
  &&\qquad\qquad\qquad\quad - \lambda \, 
  \frac{3 \, (\sigma - \mu) \, \sigma \mu}{4} \int \! \frac{d^dk}{(2 \pi)^d} \ 
  \frac{1}{[\omega_0^2 + (\gamma_0 + D_0 \, k^2)^2]^2} \ , 
\label{reg110} \\
  &&\!\!\!\!\!\!\!\!\!\! 
  \mp \omega_0 \, {\rm Im} \, \Gamma_{\pm ; \pm}^{(1)}(0,0) = 
  - \lambda \, \frac{\sigma \, \mu}{2} \int \! \frac{d^dk}{(2 \pi)^d} 
  \ \frac{1}{\omega_0^2 + (\gamma_0 + D_0 \, k^2)^2} \nonumber \\
  &&\qquad\qquad\qquad\quad + \lambda \, 
  \frac{3 \, (\sigma - \mu) \, \sigma \mu}{4} \int \! \frac{d^dk}{(2 \pi)^d} \ 
  \frac{\gamma_0 + D_0 \, k^2}{[\omega_0^2 + (\gamma_0 + D_0 \, k^2)^2]^2}  
\label{img110}  \\
  &&\qquad\qquad\qquad\quad - \lambda \, 
  \frac{(\sigma^2 - 4 \, \sigma \mu + \mu^2) \, \sigma \mu}{4} \int \!\! 
  \frac{d^dk}{(2 \pi)^d} \ \frac{1}{[\omega_0^2 + (\gamma_0 + D_0 \, k^2)^2]^2}
  \ . \nonumber
\end{eqnarray}
It is worth noting that the wavevector integrals are all of order $1 / k^4$ (or
higher inverse powers of $k$) and consequently develop ultraviolet divergences 
only in dimensions $d \geq 4$; as they should, the counter-terms have cancelled
contributions of order $1 / k^2$.
From eqs.~(\ref{rendam}) and (\ref{img110}) one immediately infers the 
fluctuation-induced damping
\begin{eqnarray}
  &&\gamma_R = \gamma_0 - \lambda \, \frac{\sigma \, \mu}{2} \int \! 
  \frac{d^dk}{(2 \pi)^d} \ \frac{1}{\omega_0^2 + (\gamma_0 + D_0 \, k^2)^2} 
  \nonumber \\
  &&\qquad\quad\ + \lambda \, \frac{3 \, (\sigma - \mu) \, \sigma \mu}{4} 
  \int \! \frac{d^dk}{(2 \pi)^d} \ 
  \frac{\gamma_0 + D_0 \, k^2}{[\omega_0^2 + (\gamma_0 + D_0 \, k^2)^2]^2} 
\label{1lpdam} \\
  &&\qquad\quad\ - \lambda \, 
  \frac{(\sigma^2 - 4 \, \sigma \mu + \mu^2) \, \sigma \mu}{4} \int \!\! 
  \frac{d^dk}{(2 \pi)^d} \ \frac{1}{[\omega_0^2 + (\gamma_0 + D_0 \, k^2)^2]^2}
  + O(\lambda^2) \ . \nonumber
\end{eqnarray}
We furthermore need
\begin{eqnarray}
  &&\!\!\!\!\!\!\!\!\!\!\!\!\!\!\!\!\!\!\!\!\!\!\!\!\!\!\!\!\!\!\!\!\! \mp 
  \omega_0 \, {\rm Im} \, \frac{\partial \, \Gamma_{\pm ; \pm}^{(1)}(0,\omega)}
  {\partial \, i \omega} \Bigg\vert_{\omega = 0} = 
  + \lambda \, \frac{\sigma - \mu}{8} \int \! \frac{d^dk}{(2 \pi)^d} \ 
  \frac{1}{\omega_0^2 + (\gamma_0 + D_0 \, k^2)^2} \nonumber \\
  &&\qquad\qquad\qquad + \lambda \, \frac{(\sigma + \mu)^2}{16} \int \! 
  \frac{d^dk}{(2 \pi)^d} \ \frac{1}{\gamma_0 + D_0 \, k^2} \ 
  \frac{1}{\omega_0^2 + (\gamma_0 + D_0 \, k^2)^2} \nonumber \\ 
  &&\qquad\qquad\qquad - \lambda \, 
  \frac{5 \, \sigma^2 - 8 \, \sigma \mu + 5 \, \mu^2}{16} \int \! 
  \frac{d^dk}{(2 \pi)^d} \ \frac{\gamma_0 + D_0 \, k^2}
  {[\omega_0^2 + (\gamma_0 + D_0 \, k^2)^2]^2} \nonumber \\
  &&\qquad\qquad\qquad - \lambda \, 
  \frac{13 \, (\sigma - \mu) \, \sigma \mu}{16} \int \! \frac{d^dk}{(2 \pi)^d} 
  \ \frac{1}{[\omega_0^2 + (\gamma_0 + D_0 \, k^2)^2]^2} \nonumber \\
  &&\qquad\qquad\qquad - \lambda \, \frac{(\sigma + \mu)^2 \, \sigma \mu}{8} 
  \int \! \frac{d^dk}{(2 \pi)^d} \ \frac{1}{\gamma_0 + D_0 \, k^2} \ 
  \frac{1}{[\omega_0^2 + (\gamma_0 + D_0 \, k^2)^2]^2} \nonumber \\
  &&\qquad\qquad\qquad + \lambda \, 
  \frac{(\sigma^2 - 4 \, \sigma \mu + \mu^2) \, \sigma \mu}{4} \int \! 
  \frac{d^dk}{(2 \pi)^d} \ \frac{\gamma_0 + D_0 \, k^2}
  {[\omega_0^2 + (\gamma_0 + D_0 \, k^2)^2]^3} \nonumber \\
  &&\qquad\qquad\qquad + \lambda \, 
  \frac{3\, (\sigma - \mu) \, \sigma^2 \mu^2}{4} \int \! \frac{d^dk}{(2 \pi)^d}
  \ \frac{1}{[\omega_0^2 + (\gamma_0 + D_0 \, k^2)^2]^3} \ ,
\label{img11w}
\end{eqnarray}
which along with (\ref{reg110}) provides us with the renormalized oscillation 
frequency (\ref{renfre})
\begin{eqnarray}
  &&\frac{\omega_R}{\omega_0} = 1 - \lambda \, \frac{\mu}{4} \int \! 
  \frac{d^dk}{(2 \pi)^d} \ \frac{1}{\omega_0^2 + (\gamma_0 + D_0 \, k^2)^2} 
  \nonumber \\
  &&\qquad\quad - \lambda \, \frac{3 \, (\sigma + \mu)^2}{16} \int \! 
  \frac{d^dk}{(2 \pi)^d} \ \frac{1}{\gamma_0 + D_0 \, k^2} \ 
  \frac{1}{\omega_0^2 + (\gamma_0 + D_0 \, k^2)^2} \nonumber \\ 
  &&\qquad\quad + \lambda \, \frac{\sigma^2 + 8 \, \sigma \mu + \mu^2}{16} 
  \int \! \frac{d^dk}{(2 \pi)^d} \ \frac{\gamma_0 + D_0 \, k^2}
  {[\omega_0^2 + (\gamma_0 + D_0 \, k^2)^2]^2} \nonumber \\
  &&\qquad\quad + \lambda \, \frac{(\sigma - \mu) \, \sigma \mu}{16} \int \! 
  \frac{d^dk}{(2 \pi)^d} \ \frac{1}{[\omega_0^2 + (\gamma_0 + D_0 \, k^2)^2]^2}
  \nonumber \\
  &&\qquad\quad + \lambda \, \frac{(\sigma + \mu)^2 \, \sigma \mu}{8} \int \! 
  \frac{d^dk}{(2 \pi)^d} \ \frac{1}{\gamma_0 + D_0 \, k^2} \ 
  \frac{1}{[\omega_0^2 + (\gamma_0 + D_0 \, k^2)^2]^2} \nonumber \\
  &&\qquad\quad - \lambda \, 
  \frac{(\sigma^2 - 4 \, \sigma \mu + \mu^2) \, \sigma \mu}{4} \int \! 
  \frac{d^dk}{(2 \pi)^d} \ \frac{\gamma_0 + D_0 \, k^2}
  {[\omega_0^2 + (\gamma_0 + D_0 \, k^2)^2]^3} \nonumber \\
  &&\qquad\quad - \lambda \, \frac{3\, (\sigma - \mu) \, \sigma^2 \mu^2}{4} 
  \int \! \frac{d^dk}{(2 \pi)^d} \ 
  \frac{1}{[\omega_0^2 + (\gamma_0 + D_0 \, k^2)^2]^3} + O(\lambda^2) \ ,
\label{1lpfre}
\end{eqnarray}
and
\begin{eqnarray}
  &&\!\!\!\!\!\!\!\!\!\!\!\!\!\!\!\!\!\!\!\!\!\!\!\!\!\!\!\!\!\!\!\!\!\! 
  \mp \, \frac{\omega_0}{D_0} \, {\rm Im} \, 
  \frac{\partial \, \Gamma_{\pm ; \pm}^{(1)}({\vec q},0)}{\partial \, q^2}
  \Bigg\vert_{{\vec q} = 0} =   
  + \lambda \, \frac{\sigma - \mu}{16} \int \! \frac{d^dk}{(2 \pi)^d} \ 
  \frac{1}{\omega_0^2 + (\gamma_0 + D_0 \, k^2)^2} \nonumber \\
  &&\qquad\qquad\qquad + \lambda \, \frac{(\sigma + \mu)^2}{16} \int \! 
  \frac{d^dk}{(2 \pi)^d} \ \frac{1}{\gamma_0 + D_0 \, k^2} \ 
  \frac{1}{\omega_0^2 + (\gamma_0 + D_0 \, k^2)^2} \nonumber \\ 
  &&\qquad\qquad\qquad - \lambda \, \frac{(\sigma + \mu)^2}{8 \, d} \int \! 
  \frac{d^dk}{(2 \pi)^d} \ \frac{D_0 \, k^2}{(\gamma_0 + D_0 \, k^2)^2} \ 
  \frac{1}{\omega_0^2 + (\gamma_0 + D_0 \, k^2)^2} \nonumber \\ 
  &&\qquad\qquad\qquad - \lambda \, \frac{3 \, (\sigma - \mu)^2}{16} \int \! 
  \frac{d^dk}{(2 \pi)^d} \ \frac{\gamma_0 + D_0 \, k^2}
  {[\omega_0^2 + (\gamma_0 + D_0 \, k^2)^2]^2} \nonumber \\
  &&\qquad\qquad\qquad + \lambda \, 
  \frac{\sigma^2 - 4 \, \sigma \mu + \mu^2}{8 \, d} \int \!
  \frac{d^dk}{(2 \pi)^d} \ \frac{D_0 \, k^2}
  {[\omega_0^2 + (\gamma_0 + D_0 \, k^2)^2]^2} \nonumber \\
  &&\qquad\qquad\qquad - \lambda \, 
  \frac{11 \, (\sigma - \mu) \, \sigma \mu}{16} \int \! \frac{d^dk}{(2 \pi)^d} 
  \ \frac{1}{[\omega_0^2 + (\gamma_0 + D_0 \, k^2)^2]^2} \nonumber \\
  &&\qquad\qquad\qquad + \lambda \, 
  \frac{3 \, (\sigma - \mu) \, \sigma \mu}{2 \, d} \int \! 
  \frac{d^dk}{(2 \pi)^d} \ \frac{D_0 \, k^2 \, (\gamma_0 + D_0 \, k^2)}
  {[\omega_0^2 + (\gamma_0 + D_0 \, k^2)^2]^3} \nonumber \\
  &&\qquad\qquad\qquad - \lambda \, \frac{(\sigma + \mu)^2 \, \sigma \mu}{16} 
  \int \! \frac{d^dk}{(2 \pi)^d} \ \frac{1}{\gamma_0 + D_0 \, k^2} \ 
  \frac{1}{[\omega_0^2 + (\gamma_0 + D_0 \, k^2)^2]^2} \nonumber \\
  &&\qquad\qquad\qquad + \lambda \, 
  \frac{(\sigma^2 - 4 \, \sigma \mu + \mu^2) \, \sigma \mu}{4} \int \! 
  \frac{d^dk}{(2 \pi)^d} \ \frac{\gamma_0 + D_0 \, k^2}
  {[\omega_0^2 + (\gamma_0 + D_0 \, k^2)^2]^3} \nonumber \\
  &&\qquad\qquad\qquad - \lambda \, 
  \frac{(\sigma^2 - 4 \, \sigma \mu + \mu^2) \, \sigma \mu}{d} \int \! 
  \frac{d^dk}{(2 \pi)^d} \ \frac{D_0 \, k^2}
  {[\omega_0^2 + (\gamma_0 + D_0 \, k^2)^2]^3} \nonumber \\
  &&\qquad\qquad\qquad + \lambda \, 
  \frac{3\, (\sigma - \mu) \, \sigma^2 \mu^2}{4} \int \! \frac{d^dk}{(2 \pi)^d}
  \ \frac{1}{[\omega_0^2 + (\gamma_0 + D_0 \, k^2)^2]^3} \nonumber \\
  &&\qquad\qquad\qquad - \lambda \, 
  \frac{3 \, (\sigma - \mu) \, \sigma^2 \mu^2}{d} \int \! 
  \frac{d^dk}{(2 \pi)^d} \ \frac{D_0 \, k^2 \, (\gamma_0 + D_0 \, k^2)}
  {[\omega_0^2 + (\gamma_0 + D_0 \, k^2)^2]^4}
\label{img11q} \\
  &&\qquad\qquad\qquad + \lambda \, 
  \frac{(\sigma^2 - 4 \, \sigma \mu + \mu^2) \, \sigma^2 \mu^2}{d} \int \! 
  \frac{d^dk}{(2 \pi)^d} \ \frac{D_0 \, k^2}
  {[\omega_0^2 + (\gamma_0 + D_0 \, k^2)^2]^4} \ , \nonumber
\end{eqnarray}
whence eq.~(\ref{rendif}) with (\ref{img11w}) at last yields the renormalized 
diffusion coefficient
\begin{eqnarray}
  &&\!\!\!\!\!\! \frac{D_R}{D_0} = 
  1 - \lambda \, \frac{\sigma - \mu}{16} \int \! \frac{d^dk}{(2 \pi)^d} \ 
  \frac{1}{\omega_0^2 + (\gamma_0 + D_0 \, k^2)^2} \nonumber \\
  &&\qquad\ - \lambda \, \frac{(\sigma + \mu)^2}{8 \, d} \int \! 
  \frac{d^dk}{(2 \pi)^d} \ \frac{D_0 \, k^2}{(\gamma_0 + D_0 \, k^2)^2} \ 
  \frac{1}{\omega_0^2 + (\gamma_0 + D_0 \, k^2)^2} \nonumber \\ 
  &&\qquad\ + \lambda \, \frac{\sigma^2 - \sigma \mu + \mu^2}{8} \int \! 
  \frac{d^dk}{(2 \pi)^d} \ \frac{\gamma_0 + D_0 \, k^2}
  {[\omega_0^2 + (\gamma_0 + D_0 \, k^2)^2]^2} \nonumber \\
  &&\qquad\ + \lambda \, 
  \frac{\sigma^2 - 4 \, \sigma \mu + \mu^2}{8 \, d} \int \!
  \frac{d^dk}{(2 \pi)^d} \ \frac{D_0 \, k^2}
  {[\omega_0^2 + (\gamma_0 + D_0 \, k^2)^2]^2} \nonumber \\
  &&\qquad\ + \lambda \, \frac{(\sigma - \mu) \, \sigma \mu}{8} \int \! 
  \frac{d^dk}{(2 \pi)^d} \ \frac{1}{[\omega_0^2 + (\gamma_0 + D_0 \, k^2)^2]^2}
  \nonumber \\
  &&\qquad\ + \lambda \, 
  \frac{3 \, (\sigma - \mu) \, \sigma \mu}{2 \, d} \int \! 
  \frac{d^dk}{(2 \pi)^d} \ \frac{D_0 \, k^2 \, (\gamma_0 + D_0 \, k^2)}
  {[\omega_0^2 + (\gamma_0 + D_0 \, k^2)^2]^3} \nonumber \\
  &&\qquad\ + \lambda \, \frac{(\sigma + \mu)^2 \, \sigma \mu}{16} 
  \int \! \frac{d^dk}{(2 \pi)^d} \ \frac{1}{\gamma_0 + D_0 \, k^2} \ 
  \frac{1}{[\omega_0^2 + (\gamma_0 + D_0 \, k^2)^2]^2} \nonumber \\
  &&\qquad\ - \lambda \, 
  \frac{(\sigma^2 - 4 \, \sigma \mu + \mu^2) \, \sigma \mu}{d} \int \! 
  \frac{d^dk}{(2 \pi)^d} \ \frac{D_0 \, k^2}
  {[\omega_0^2 + (\gamma_0 + D_0 \, k^2)^2]^3} \nonumber \\
  &&\qquad\ - \lambda \, 
  \frac{3 \, (\sigma - \mu) \, \sigma^2 \mu^2}{d} \int \! 
  \frac{d^dk}{(2 \pi)^d} \ \frac{D_0 \, k^2 \, (\gamma_0 + D_0 \, k^2)}
  {[\omega_0^2 + (\gamma_0 + D_0 \, k^2)^2]^4}
\label{1lpdif} \\
  &&\qquad\ + \lambda \, 
  \frac{(\sigma^2 - 4 \, \sigma \mu + \mu^2) \, \sigma^2 \mu^2}{d} \int \! 
  \frac{d^dk}{(2 \pi)^d} \ \frac{D_0 \, k^2}
  {[\omega_0^2 + (\gamma_0 + D_0 \, k^2)^2]^4} + O(\lambda^2) \ . \nonumber 
\end{eqnarray}

Finally, the wavevector integrals for the fluctuation corrections need to be
carried out, as sketched in \ref{inttab}.

\section{Wavevector integrals}
\label{inttab}

The required integrals are of the following form, and readily evaluated
(where convergent in the ultraviolet) by means of Euler's Gamma function:
\begin{eqnarray}
  &&\int \! \frac{d^dk}{(2 \pi)^d} \, 
  \frac{k^{2 \sigma}}{\left( \tau + k^2 \right)^s} = 
  \frac{1}{2^{d-1} \, \pi^{d/2} \, \Gamma(d/2)} \int_0^\infty \! 
  \frac{k^{d - 1 + 2 \sigma}}{(\tau + k^2)^s} \, dk \nonumber \\
  &&\qquad\qquad\qquad\quad\ \ 
  = \frac{\Gamma(\sigma + d/2) \, \Gamma(s - \sigma - d/2)}
  {2^d \, \pi^{d/2} \, \Gamma(d/2) \, \Gamma(s)} \ \tau^{\sigma - s + d/2} \ .
\label{funint}
\end{eqnarray}
This immediately yields the basic integrals
\begin{eqnarray}
  &&\int \! \frac{d^dk}{(2 \pi)^d} \, 
  \frac{1}{\omega_0^2 + (\gamma_0 + D_0 \, k^2)^2} = 
  - \frac{1}{\omega_0 \, D_0} \, {\rm Im} \int \! \frac{d^dk}{(2 \pi)^d} \, 
  \frac{1}{k^2 + (\gamma_0 + i \omega_0) / D_0} \nonumber \\
  &&\qquad\qquad\qquad\qquad = - \frac{\Gamma(1 - d/2)}{2^d \, \pi^{d/2}} \, 
  \frac{\omega_0^{-2+d/2}}{D_0^{d/2}} \, 
  {\rm Im} \, \Bigl( \frac{\gamma_0}{\omega_0} + i \Bigr)^{-1+d/2} ,
\label{kint1} \\
  &&\int \! \frac{d^dk}{(2 \pi)^d} \, 
  \frac{D_0 \, k^2}{\omega_0^2 + (\gamma_0 + D_0 k^2)^2} = 
  - \frac{1}{\omega_0} \, {\rm Im} \int \! \frac{d^dk}{(2 \pi)^d} \, 
  \frac{k^2}{k^2 + (\gamma_0 + i \omega_0) / D_0} \nonumber \\
  &&\qquad\qquad\qquad\qquad = \frac{\Gamma(1 - d/2)}{2^d \, \pi^{d/2}} \, 
  \frac{\omega_0^{-1+d/2}}{D_0^{d/2}} \, 
  {\rm Im} \, \Bigl( \frac{\gamma_0}{\omega_0} + i \Bigr)^{d/2} ,
\label{kint2} \\  
  &&\int \! \frac{d^dk}{(2 \pi)^d} \, 
  \frac{\gamma_0 + D_0 \, k^2}{\omega_0^2 + (\gamma_0 + D_0 k^2)^2} = 
  \frac{1}{D_0} \, {\rm Re} \int \! \frac{d^dk}{(2 \pi)^d} \, 
  \frac{1}{k^2 + (\gamma_0 + i \omega_0) / D_0} \nonumber \\
  &&\qquad\qquad\qquad\qquad = \frac{\Gamma(1 - d/2)}{2^d \, \pi^{d/2}} \, 
  \frac{\omega_0^{-1+d/2}}{D_0^{d/2}} \, 
  {\rm Re} \, \Bigl( \frac{\gamma_0}{\omega_0} + i \Bigr)^{-1+d/2} .
\label{kint3}
\end{eqnarray}
This last result also follows from the sum of eqs.~(\ref{kint1}) and 
(\ref{kint2}), if one observes that
$$ {{\rm Re} \choose {\rm Im}}\, \Bigl( \frac{\gamma_0}{\omega_0} + i \Bigr)^k
  = \frac{\gamma_0}{\omega_0} \, {{\rm Re} \choose {\rm Im}} \, \Bigl( 
  \frac{\gamma_0}{\omega_0} + i \Bigr)^{k-1} + {- {\rm Im} \choose {\rm Re}} \,
  \Bigl( \frac{\gamma_0}{\omega_0} + i \Bigr)^{k-1} \ . $$
Next, decomposition into partial fractions gives
\begin{eqnarray}
  &&\int \! \frac{d^dk}{(2 \pi)^d} \, \frac{1}{\gamma_0 + D_0 \, k^2} \, 
  \frac{1}{\omega_0^2 + (\gamma_0 + D_0 k^2)^2} \nonumber \\
  &&\qquad\quad = \frac{1}{\omega_0^2 \, D_0} \int \! \frac{d^dk}{(2 \pi)^d} \,
  \Biggl( \frac{1}{k^2 + \gamma_0 / D_0} - \, {\rm Re} \, 
  \frac{1}{k^2 + (\gamma_0 + i \omega_0) / D_0} \Biggr) \nonumber \\
  &&\qquad\quad = \frac{\Gamma(1 - d/2)}{2^d \, \pi^{d/2}} \, 
  \frac{\omega_0^{-3+d/2}}{D_0^{d/2}} \, 
  \Biggl[ \Bigl( \frac{\gamma_0}{\omega_0} \Bigr)^{-1+d/2} 
  - {\rm Re} \, \Bigl( \frac{\gamma_0}{\omega_0} + i \Bigr)^{-1+d/2} \Biggr] ,
\label{kint4} \\
  &&\int \! \frac{d^dk}{(2 \pi)^d} \, 
  \frac{D_0 \, k^2}{(\gamma_0 + D_0 \, k^2)^2} \, 
  \frac{1}{\omega_0^2 + (\gamma_0 + D_0 k^2)^2} \nonumber \\
  &&\qquad\quad = \frac{1}{\omega_0^2 \, D_0} \int \! \frac{d^dk}{(2 \pi)^d} \,
  \Biggl( \frac{k^2}{(k^2 + \gamma_0 / D_0)^2} - \, 
  \frac{k^2}{(k^2 + \gamma_0 / D_0)^2 + \omega_0^2 / D_0^2} \Biggr) \nonumber\\
  &&\qquad\quad = \frac{\Gamma(1 - d/2)}{2^d \, \pi^{d/2}} \, 
  \frac{\omega_0^{-3+d/2}}{D_0^{d/2}} \, \Biggl[ \, \frac{d}{2} \, 
  \Bigl( \frac{\gamma_0}{\omega_0} \Bigr)^{-1+d/2} 
  - {\rm Im} \, \Bigl( \frac{\gamma_0}{\omega_0} + i \Bigr)^{d/2} \Biggr] .
\label{kint5} 
\end{eqnarray}  

Taking derivatives with respect to the parameters $\gamma_0$ and/or $\omega_0$
one then obtains:
\begin{eqnarray}
  &&\int \! \frac{d^dk}{(2 \pi)^d} \, 
  \frac{\gamma_0 + D_0 \, k^2}{[\omega_0^2 + (\gamma_0 + D_0 k^2)^2]^2} = 
  - \, \frac{1}{2} \, \frac{\partial}{\partial \gamma_0} 
  \int \! \frac{d^dk}{(2 \pi)^d} \, 
  \frac{1}{\omega_0^2 + (\gamma_0 + D_0 \, k^2)^2} \nonumber \\
  &&\qquad\qquad\qquad\qquad = - \, \frac{\Gamma(2 - d/2)}{2^{d+1}\, \pi^{d/2}}
  \, \frac{\omega_0^{-3+d/2}}{D_0^{d/2}} \, 
  {\rm Im} \, \Bigl( \frac{\gamma_0}{\omega_0} + i \Bigr)^{-2+d/2} ,
\label{dint1} \\
  &&\int \! \frac{d^dk}{(2 \pi)^d} \, 
  \frac{D_0 \, k^2}{[\omega_0^2 + (\gamma_0 + D_0 k^2)^2]^2} = 
  - \, \frac{1}{2 \, \omega_0} \, \frac{\partial}{\partial \omega_0} \int \! 
  \frac{d^dk}{(2 \pi)^d} \, 
  \frac{D_0 \, k^2}{\omega_0^2 + (\gamma_0 + D_0 k^2)^2} \nonumber \\
  &&\quad = \frac{\Gamma(1 - d/2)}{2^{d+1} \, \pi^{d/2}} \, 
  \frac{\omega_0^{-3+d/2}}{D_0^{d/2}} \, \Biggl[ {\rm Im} \, \Bigl( 
  \frac{\gamma_0}{\omega_0} + i \Bigr)^{d/2} - \frac{d}{2} \, {\rm Re} \, 
  \Bigl( \frac{\gamma_0}{\omega_0} + i \Bigr)^{-1+d/2} \Biggr] ,
\label{dint2} \\
  &&\int \!\! \frac{d^dk}{(2 \pi)^d} \, 
  \frac{1}{[\omega_0^2 + (\gamma_0 + D_0 k^2)^2]^2} = 
  - \, \frac{\Gamma(1 - d/2)}{2^{d+1} \, \pi^{d/2}} \, 
  \frac{\omega_0^{-4+d/2}}{D_0^{d/2}} \, 
  {\rm Im} \, \Bigl( \frac{\gamma_0}{\omega_0} + i \Bigr)^{-1+d/2} \nonumber \\
  &&\qquad\qquad\qquad\qquad\quad - \frac{\Gamma(2 - d/2)}{2^{d+1}\,\pi^{d/2}}
  \, \frac{\omega_0^{-4+d/2}}{D_0^{d/2}} \, 
  {\rm Re} \, \Bigl( \frac{\gamma_0}{\omega_0} + i \Bigr)^{-2+d/2} ,
\label{dint3} \\
  &&\!\! \int \!\! \frac{d^dk}{(2 \pi)^d} \, \frac{D_0 \, k^2 \,
  (\gamma_0 + D_0 \, k^2)}{[\omega_0^2 + (\gamma_0 + D_0 k^2)^2]^3} = 
  - \frac{d \, \Gamma(1 - d/2)}{2^{d+4} \, \pi^{d/2}} \,
  \frac{\omega_0^{-4+d/2}}{D_0^{d/2}} \, 
  {\rm Im} \Bigl( \frac{\gamma_0}{\omega_0} + i \Bigr)^{-1+d/2} \nonumber \\
  &&\qquad\qquad\qquad\qquad\quad 
  - \frac{d \, \Gamma(2 - d/2)}{2^{d+4} \, \pi^{d/2}} \, 
  \frac{\omega_0^{-4+d/2}}{D_0^{d/2}} \, 
  {\rm Re} \, \Bigl( \frac{\gamma_0}{\omega_0} + i \Bigr)^{-2+d/2} ,
\label{dint4} \\
  &&\int \! \frac{d^dk}{(2 \pi)^d} \, \frac{1}{\gamma_0 + D_0 \, k^2} \, 
  \frac{1}{[\omega_0^2 + (\gamma_0 + D_0 k^2)^2]^2} \nonumber \\
  &&\qquad\qquad\qquad = \frac{\Gamma(1 - d/2)}{2^d \, \pi^{d/2}} \, 
  \frac{\omega_0^{-5+d/2}}{D_0^{d/2}} \, 
  \Biggl[ \Bigl( \frac{\gamma_0}{\omega_0} \Bigr)^{-1+d/2} 
  - {\rm Re} \, \Bigl( \frac{\gamma_0}{\omega_0} + i \Bigr)^{-1+d/2} \Biggr] 
  \nonumber \\
  &&\qquad\qquad\qquad\qquad\quad + \frac{\Gamma(2 - d/2)}{2^{d+1}\, \pi^{d/2}}
  \, \frac{\omega_0^{-5+d/2}}{D_0^{d/2}} \, 
  {\rm Im} \, \Bigl( \frac{\gamma_0}{\omega_0} + i \Bigr)^{-2+d/2} ,
\label{dint5} \\
  &&\int \! \frac{d^dk}{(2 \pi)^d} \, 
  \frac{\gamma_0 + D_0 \, k^2}{[\omega_0^2 + (\gamma_0 + D_0 k^2)^2]^3} = 
  - \, \frac{\Gamma(2 - d/2)}{2^{d+3} \, \pi^{d/2}} \,
  \frac{\omega_0^{-5+d/2}}{D_0^{d/2}} \, 
  {\rm Im} \Bigl( \frac{\gamma_0}{\omega_0} + i \Bigr)^{-2+d/2} \nonumber \\
  &&\qquad\qquad\qquad\qquad\quad 
  - \, \frac{\Gamma(3 - d/2)}{2^{d+3} \, \pi^{d/2}} \, 
  \frac{\omega_0^{-5+d/2}}{D_0^{d/2}} \, 
  {\rm Re} \, \Bigl( \frac{\gamma_0}{\omega_0} + i \Bigr)^{-3+d/2} ,
\label{dint6} \\
  &&\int \! \frac{d^dk}{(2 \pi)^d} \, 
  \frac{D_0 \, k^2}{[\omega_0^2 + (\gamma_0 + D_0 k^2)^2]^3} \nonumber \\
  &&\qquad\quad\ = \frac{3 \, \Gamma(1 - d/2)}{2^{d+3} \, \pi^{d/2}} \, 
  \frac{\omega_0^{-5+d/2}}{D_0^{d/2}} \, \Biggl[ 
  {\rm Im} \, \Bigl( \frac{\gamma_0}{\omega_0} + i \Bigr)^{d/2}\! - \frac{d}{2}
  \, {\rm Re} \, \Bigl( \frac{\gamma_0}{\omega_0} + i \Bigr)^{-1+d/2} 
  \Biggr] \nonumber \\
  &&\qquad\qquad\qquad\qquad\quad 
  + \frac{d \, \Gamma(2 - d/2)}{2^{d+4} \, \pi^{d/2}}
  \, \frac{\omega_0^{-5+d/2}}{D_0^{d/2}} \, 
  {\rm Im} \, \Bigl( \frac{\gamma_0}{\omega_0} + i \Bigr)^{-2+d/2} ,
\label{dint7} \\
  &&\!\! \int \!\! \frac{d^dk}{(2 \pi)^d} \, 
  \frac{1}{[\omega_0^2 + (\gamma_0 + D_0 k^2)^2]^3} = 
  - \frac{3 \, \Gamma(1 - d/2)}{2^{d+3} \, \pi^{d/2}} \, 
  \frac{\omega_0^{-6+d/2}}{D_0^{d/2}} \, 
  {\rm Im} \, \Bigl( \frac{\gamma_0}{\omega_0} + i \Bigr)^{-1+d/2} \nonumber \\
  &&\qquad\qquad\qquad\qquad\quad 
  - \, \frac{3 \, \Gamma(2 - d/2)}{2^{d+3} \, \pi^{d/2}} \, 
  \frac{\omega_0^{-6+d/2}}{D_0^{d/2}} \, 
  {\rm Re} \, \Bigl( \frac{\gamma_0}{\omega_0} + i \Bigr)^{-2+d/2} \nonumber \\
  &&\qquad\qquad\qquad\qquad\quad
  + \frac{\Gamma(3 - d/2)}{2^{d+3} \, \pi^{d/2}} \, 
  \frac{\omega_0^{-6+d/2}}{D_0^{d/2}} \, 
  {\rm Im} \Bigl( \frac{\gamma_0}{\omega_0} + i \Bigr)^{-3+d/2} ,
\label{dint8} \\
  &&\!\!\! \int \!\! \frac{d^dk}{(2 \pi)^d} \, \frac{D_0 \, k^2 \,
  (\gamma_0 + D_0 \, k^2)}{[\omega_0^2 + (\gamma_0 + D_0 k^2)^2]^4} =   
  - \frac{d \, \Gamma(1 - d/2)}{2^{d+5} \, \pi^{d/2}} \, 
  \frac{\omega_0^{-6+d/2}}{D_0^{d/2}} \, 
  {\rm Im} \, \Bigl( \frac{\gamma_0}{\omega_0} + i \Bigr)^{-1+d/2} \nonumber \\
  &&\qquad\qquad\qquad\qquad\quad 
  - \, \frac{d \, \Gamma(2 - d/2)}{2^{d+5} \, \pi^{d/2}} \, 
  \frac{\omega_0^{-6+d/2}}{D_0^{d/2}} \, 
  {\rm Re} \, \Bigl( \frac{\gamma_0}{\omega_0} + i \Bigr)^{-2+d/2} \nonumber \\
  &&\qquad\qquad\qquad\qquad\quad
  + \frac{d \, \Gamma(3 - d/2)}{3 \cdot 2^{d+5} \, \pi^{d/2}} \, 
  \frac{\omega_0^{-6+d/2}}{D_0^{d/2}} \, 
  {\rm Im} \Bigl( \frac{\gamma_0}{\omega_0} + i \Bigr)^{-3+d/2} ,
\label{dint9} \\
  &&\int \! \frac{d^dk}{(2 \pi)^d} \, 
  \frac{D_0 \, k^2}{[\omega_0^2 + (\gamma_0 + D_0 k^2)^2]^4} \nonumber \\
  &&\qquad\quad\ = \frac{5 \, \Gamma(1 - d/2)}{2^{d+4} \, \pi^{d/2}} \, 
  \frac{\omega_0^{-7+d/2}}{D_0^{d/2}} \, \Biggl[ 
  {\rm Im} \, \Bigl( \frac{\gamma_0}{\omega_0} + i \Bigr)^{d/2}\! - \frac{d}{2}
  \, {\rm Re} \, \Bigl( \frac{\gamma_0}{\omega_0} + i \Bigr)^{-1+d/2} 
  \Biggr] \nonumber \\
  &&\qquad\qquad\qquad\qquad\quad 
  + \frac{d \, \Gamma(2 - d/2)}{2^{d+4} \, \pi^{d/2}}
  \, \frac{\omega_0^{-7+d/2}}{D_0^{d/2}} \, 
  {\rm Im} \, \Bigl( \frac{\gamma_0}{\omega_0} + i \Bigr)^{-2+d/2} \nonumber \\
  &&\qquad\qquad\qquad\qquad\quad 
  + \frac{d \, \Gamma(3 - d/2)}{3 \cdot 2^{d+5} \, \pi^{d/2}}
  \, \frac{\omega_0^{-7+d/2}}{D_0^{d/2}} \, 
  {\rm Re} \, \Bigl( \frac{\gamma_0}{\omega_0} + i \Bigr)^{-3+d/2} .
\label{dint0} 
\end{eqnarray}

For explicit evaluation at $d = 2$, the Gamma function $\Gamma(1 - d/2)$ 
diverges, but its poles in the expressions for the renormalized oscillation 
parameters are all cancelled, as can be checked by setting 
$d = 2 - \varepsilon$, and carefully taking the limit $\varepsilon \to 0$.
Indeed, the singularities in two dimensions are eliminated by the counter-terms
$A_c$ and $B_c$.
At $d = 4$, ultraviolet divergences appear, which must be regularized by a
cut-off $\Lambda$ in momentum space that originates from the underlying 
lattice; e.g., $\Lambda = 2 \pi / a_0$ in a hypercubic lattice with lattice
constant $a_0$.
In the above integral listing, these ultraviolet singularities emerge as poles
in $\epsilon = 4 - d$.
For example, the logarithmic cutoff dependence 
$\frac14 \, \ln (1 + \Lambda^4 D_0^2 / \omega_0^2)$ is represented in
dimensional regularization by 
$\Gamma(1 + \epsilon / 2) / \epsilon (1 - \epsilon / 2)$.

\end{document}